\documentclass[preprint]{aastex61}

\newcommand\PDF{P}
\newcommand\CCDF{F}
\newcommand\xmin{x_{\rm min}}
\received{April 12, 2022}
\revised{November 4, 2022}
\accepted{November 14, 2022}

\submitjournal{ApJ}

\shorttitle{Magnetic Flux Distribution of Sunspots}
\shortauthors{Sakurai \& Toriumi}

\begin{document}

\title{Probability Distribution Functions of Sunspot Magnetic Flux}

\correspondingauthor{Takashi Sakurai}
\email{takashi.sakurai@nao.ac.jp}

\author[0000-0002-6019-5167]{Takashi Sakurai}
\affiliation{National Astronomical Observatory of Japan, 2-21-1 Osawa, Mitaka, Tokyo 181-8588, Japan}
\affiliation{Earth-Life Science Institute, Tokyo Institute of Technology, 2-12-1 Ookayama, Meguro, Tokyo 152-8550, Japan}

\author[0000-0002-1276-2403]{Shin Toriumi}
\affiliation{Institute of Space and Astronautical Science, Japan Aerospace Exploration Agency, Sagamihara, Kanagawa 229-5210, Japan}




\begin{abstract}
We have investigated the probability distributions of sunspot area and magnetic flux by using the data from Royal Greenwich Observatory and USAF/NOAA. We have constructed a sample of 2995 regions with maximum-development areas $\ge$ 500 MSH (millionths of solar hemisphere), covering 146.7 years (1874--2020).
The data were fitted by a power-law distribution and four two-parameter distributions (tapered power-law, gamma, lognormal, and Weibull distributions). 
The power-law model was unfavorable compared to the four models in terms of AIC, and was not acceptable by the classical Kolmogorov-Smirnov test. The lognormal and Weibull distributions were excluded because their behavior extended to smaller regions ($S \ll 500$ MSH) do not connect to the previously published results. Therefore, our choices were tapered power-law and gamma distributions.
The power-law portion of the tapered power-law and gamma distributions was found to have a power exponent of 1.35--1.9.
Due to the exponential fall-off of these distributions, the expected frequencies of large sunspots are low.
The largest sunspot group observed had an area of 6132 MSH, and the frequency of sunspots larger than $10^4$ MSH was estimated to be every 3 -- 8 $\times 10^4$ years.
We also have estimated the distributions of the Sun-as-a-star total sunspot areas. The largest total area covered by sunspots in the record was 1.67 \% of the visible disk, and can be up to 2.7 \% by artificially increasing the lifetimes of large sunspots in an area evolution model. These values are still smaller than those found on active Sun-like stars.
\end{abstract}





\section{Introduction}
\label{sec:intro}
Sunspots represent a variety of magnetic activities of the Sun \citep{sol03,van15}. It is thought that the dynamo mechanism in the solar interior intensifies and transports magnetic flux to the surface and eventually builds up active regions (ARs) including sunspots \citep{par55}. Even after four centuries of continuous observations, the sunspots still maintain important positions in the investigation of the dynamo processes in the Sun.

\begin{figure*}[htbp]
\begin{center}
\includegraphics[width=80mm]{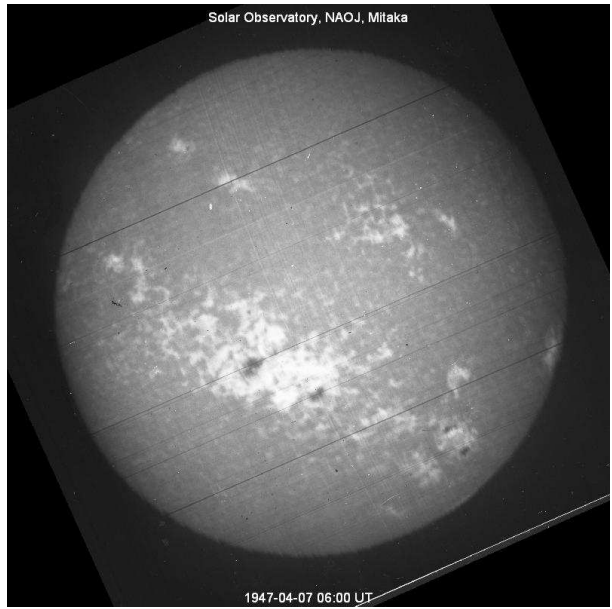}
\end{center}
\caption{
Ca K-line spectroheliogram of 1947 April 7 taken at Mitaka, Tokyo; celestial north is up, west to the right. The big active region near the disk center is RGO region 14886 which showed a foreshortening-corrected area of 6132 MSH, the largest on the record, on April 8.
\label{fig:region1947}
}
\end{figure*}

Another importance of sunspots is their close relationship with flare activity \citep{pri02,shi11,tor19}. Statistical investigations have revealed that greater flares emanate from larger ARs \citep{sam00}. This may be natural since larger ARs harbor more magnetic flux and thus more magnetic free energy available. The largest observed sunspot group since the late nineteenth century was the one in 1947 April, the largest area of which was 6132 MSH (millionths of solar hemisphere; 1 MSH = $3.04 \times 10^6$ km$^2$) or about 1.2 \% of the visible solar disk (Figure \ref{fig:region1947}). While this region was not flare-active, another giant sunspot in 1946 July caused larger flares with geomagnetic disturbances \citep{tor17}. The formation mechanism of such great ARs is an interesting issue to be resolved.

The existence of spots is also known for other stars \citep{ber05,str09}. One of the largest starspots reported thus far was from a K0 giant XX Tri (HD 12545), which covered about 20\% of the entire stellar surface \citep{str99}. It was found that even solar-type stars producing the so-called superflares \citep{sch00,mae12} host starspots much larger than the solar ones \citep[up to $\sim 10\%$ of the stellar hemisphere;][]{not13,not15}. Therefore, the discussion of superflares on the Sun is closely related to the question of the production of super-large sunspots.

The key quantity we study in this paper is the probability distribution function of the area $S$ or magnetic flux $\Phi$ of sunspots. \citet{bog88} analyzed the Mt.~Wilson white-light observations (1917--1982) of sunspots and showed that the sunspot umbral areas [1.5--141 MSH; the corresponding total sunspot areas would be about five times of these \citep{sol03}] follow the lognormal distribution. \citet{hat08} obtained the same conclusion using the data from the Royal Greenwich Observatory (RGO) and the United States Air Force [USAF; data were compiled and distributed by the National Oceanic and Atmospheric Administration (NOAA)] covering the period 1874--2007 (sunspot areas larger than 35 MSH). \citet{bau05} studied the RGO data (1874--1976) of sunspots with areas $\geq 60$ MSH, by making a distinction between a snapshot distribution and a maximum area distribution; the former is derived from daily data \citep[e.g.][]{bog88, hat08} while the latter is derived by following the time evolution of individual regions and by recording their maximum areas. They found that the two distributions are fitted by lognormal distributions that have similar parameter values. This property was also mentioned in \citet{hat08}.

The fall-off of the lognormal distribution toward small sunspot areas may be because smaller magnetic concentrations tend to lose their darkness and eventually end up with small flux tubes brighter than the surroundings \citep{zwa87}. On the other hand, the fall-off of the lognormal distribution toward large sunspot areas was not paid much attention. In this context, \citet[][Figure 6]{gop18} applied the Weibull distribution \citep{wei39}, which is often used to describe the failure rates of industrial products, to the RGO and USAF/NOAA sunspot area data (maximum-area distribution). The tail of the observed distribution was fitted equally well by a power law and Weibull, and the latter gives more rapid decline and less frequent appearance of extremely large regions.

Since sunspots are made (i.e. obtain their darkness) because of their magnetic fields by inhibition of convective heat transport \citep[e.g.][]{spr77}, the probability distribution of magnetic flux in magnetic structures (including sunspots) is an equally important quantity. \citet{khz93} analyzed the magnetograms obtained at the US National Solar Observatory at Kitt Peak (NSO/Kitt Peak) in the period 1975--1986 and derived the distribution of magnetic flux at the maximum development of individual active regions. For regions with areas larger than 121 MSH the magnetic flux emergence rate was approximated by a power law. \citet{sch94} gave a more specific fitting equation with a power-law exponent of about 2.0 [probability $\propto \Phi^{-\alpha}$ with $\alpha \simeq 2$; in this article $\alpha (>0)$ is called the power-law exponent]. \citet{hst03} investigated the emergence rates of ephemeral regions (small-scale bipolar magnetic field patches) using the data from the Michelson Doppler Imager onboard the Solar and Heliospheric Observatory \citep[SOHO/MDI;][]{sch95} taken between 1996 and 2001. They found that the distribution looked like an exponential function. \citet{tho11} analyzed the emergence rates of small-scale magnetic field patches using the data from the Solar Optical Telescope onboard the Hinode satellite \citep[Hinode/SOT;][]{tsu08} and obtained a power-law formula extending all the way up to the active-region scales, with a power-law exponent of 2.69. All of these results are summarized and discussed later in Section \ref{sec:comparison} and in Figure \ref{fig:comparison}. \citet{bal09}, \citet{mun15}, and \citet{man20} give detailed accounts on the issues on sunspot area calibration. \citet{mun15} also tried several models (power law, lognormal, Weibul, exponential, and their combinations) to fit the area distributions.

In this paper, we will use the data from RGO (1874--1976) and USAF/NOAA (1977 - 2020) and derive the maximum areas of individual regions (Section \ref{sec:data}). Recurrent regions are counted only once when their areas reach the maximum. In order to compare these with magnetic field observations of active regions and smaller magnetic patches, we will convert the sunspot areas to magnetic flux values (Section \ref{sec:area_vs_flux}). Our primary interest here is whether the sunspot area or magnetic flux distribution extends to large values in the form of a power law or is tapered off, to address whether the Sun may have super-large sunspots like in super-flare stars. Therefore, we limit the data of sunspots with areas 500 MSH or larger, and try to fit the data with five kinds of distribution functions; power law, tapered power law, gamma, lognormal, and Weibull distributions (Section \ref{sec:fitting_models}). Statistical examinations (Section \ref{sec:fitting_results}) and comparison with previously published results (Section \ref{sec:comparison}) show preference on the tapered power law and gamma distributions. Using the obtained results, we can predict the expected frequencies of super-large sunspots. In Section \ref{sec:model}, we will adopt a simple time-evolution model of sunspot areas and examine the effects of assumptions we made in our analysis. Particularly we can estimate the snapshot (instantaneous) distribution of sunspot areas (Section \ref{sec:instantaneous}) and also a distribution of Sun-as-a-star total sunspot areas (Section \ref{sec:whole-Sun}), and will discuss their implications on super-large stellar spots (Section \ref{sec:summary}).

\section{Data}
\label{sec:data}
\subsection{Data Sources}
\label{sec:data_sources}
%
The Greenwich Photoheliographic Results (GPR) in PDF are available through the SAO/NASA Astrophysics Data System (ADS), NOAA%
\footnote{ftp://ftp.ngdc.noaa.gov/STP/SOLAR\_DATA/SOLAR\_OBSERVATION/GREENWICH/}, 	
and the UK Solar System Data Center%
\footnote{http://www.ukssdc.ac.uk/wdcc1/RGOPHR/}.	
The digitized data of sunspot areas recorded in GPR (1874 April -- 1976) are provided at several sites%
\footnote{ftp://ftp.ngdc.noaa.gov/STP/SOLAR\_DATA/SUNSPOT\_REGIONS/Greenwich/}$^{,}$
\footnote{http://fenyi.solarobs.csfk.mta.hu/GPR/}$^{,}$
\footnote{http://solarcyclescience.com/activeregions.html}.	
Systematic observation and data collection of sunspots at RGO started on 1874 April 17 \citep{chr1907}. The correction for foreshortening was explained to have been applied to regions of angular distance within $80^\circ$ from the disk center \citep{rgo55}, but regions beyond were occasionally recorded. From these data sets we have picked up regions whose foreshortening-corrected maximum-development areas ($S_{\rm md}$) are 500 MSH or larger. If the corrected area $S$ of a region was monotonically increasing or decreasing on the visible hemisphere, we defined the maximum value of $S$ as the maximum-development area $S_{\rm md}$, although the true maximum took place on the back-side of the Sun. The effects of this assumption will be estimated in Section \ref{sec:model}. For recurrent regions, we only retained the maximum $S$ over all their disk passages because they were regarded as generated from the identical magnetic flux tubes generated by the dynamo. For this we had to identify recurrent regions, and such lists are available in \citet{mau1909} (1874--1906), and ``Catalogue of Recurrent Groups of Sun Spots'' (1910--1955), ``Ledger I: Recurrent Groups'' (1916--1955), and ``General Catalog of Groups of Sunspots'' (1956--1976) sections of GPR.

The data after the cessation of RGO solar observations in 1976 were taken from USAF/NOAA (1977--2020)$^5$.
As no convenient lists are available to identify recurrent regions, we did this manually by relying on the following data: H$\alpha$ synoptic charts in Solar Geophysical Data (SGD)%
\footnote{ftp://ftp.ngdc.noaa.gov/STP/SOLAR\_DATA/SGD\_PDFversion/} (1977--1989),	
NOAA Report of Solar and Geophysical Activity (RSGA)%
\footnote{ftp://ftp.swpc.noaa.gov/pub/warehouse/} (1990--2000),	
NOAA Weekly Preliminary Report and Forecast of SGD (Weekly PRF)$^7$
(2001--2009), and
Debrecen Photoheliographic Data (DPD)%
\footnote{http://fenyi.solarobs.csfk.mta.hu/DPD/} (2010--2020).	
We have picked up regions with the foreshortening-corrected maximum-development area exceeding 500/1.20 = 415 MSH (see below). The distribution of angular distances from the disk center did not show a clear decline toward the limb and continued up to $90^\circ$, implying that the foreshortening correction was applied to all the USAF/NOAA data.

In the end we have selected 2995 regions, 2175 from RGO and 820 from USAF/NOAA data, covering 1874 April to 2020 December, 146.7 years (Table \ref{tab:data}). Before obtaining the final list, we have removed 363 and 101 regions from RGO and USAF/NOAA data as they were members of recurrent regions. The data fully cover 13 solar cycles, from Cycle 12 (1878 December -- 1890 March) to Cycle 24 (2008 December -- 2019 December).

\begin{table}[htbp]
\begin{center}
\caption{
Data
}
\label{tab:data}
\begin{tabular}{rrrrrrr}
\hline
\hline
No. & \multicolumn{1}{c}{Date}  & Data source & Region number & RGO area & Original area$^a$ & Magnetic flux \cr
    &           &          &              & [MSH]    & [MSH]         & [Mx] \cr
\hline
1  & 1947-04-08 & RGO\phantom{A} & 14886  & 6132.0  & 6132  & 3.18E+23 \\
2  & 1946-02-07 & RGO\phantom{A} & 14417  & 5202.0  & 5202  & 2.69E+23 \\
3  & 1951-05-19 & RGO\phantom{A} & 16763  & 4865.0  & 4865  & 2.51E+23 \\
4  & 1946-07-29 & RGO\phantom{A} & 14585  & 4720.0  & 4720  & 2.44E+23 \\
5  & 1989-03-18 & NOAA           &  5395  & 4320.0  & 3600  & 2.23E+23 \\
6  & 1982-06-15 & NOAA           &  3776  & 3720.0  & 3100  & 1.92E+23 \\
7  & 1926-01-19 & RGO\phantom{A} &  9861  & 3716.0  & 3716  & 1.92E+23 \\
8  & 1989-09-04 & NOAA           &  5669  & 3696.0  & 3080  & 1.91E+23 \\
9  & 1990-11-19 & NOAA           &  6368  & 3696.0  & 3080  & 1.91E+23 \\
10 & 1938-01-21 & RGO\phantom{A} & 12673  & 3627.0  & 3627  & 1.87e+23 \\
\hline
\hline
\multicolumn{7}{l}{$^a$: NOAA area before converted to the RGO scale.}
\end{tabular}
\end{center}
\tablecomments{Table \ref{tab:data} is published in its entirety in the machine-readable format. Only the top ten regions are shown here to explain the format of the table.}
\end{table}

\subsection{Sunspot Area vs. Magnetic Flux}
\label{sec:area_vs_flux}
%
In order to relate the sunspot area and the total radial unsigned magnetic flux (including the flux outside of the sunspots), we used the SHARP data series \citep{bob14} of the Helioseismic and Magnetic Imager (HMI; \citealt{sche12,scho12}) aboard the Solar Dynamics Observatory (SDO; \citealt{pes12}). We picked up all available CEA (cylindrical equal area)-remapped definitive SHARP data from 2010 to 2015 that contained the sunspots of NOAA areas $\ge 500\ {\rm MSH}$ within 45$^{\circ}$ from the disk center. In total, 137 patches were collected (one patch per region per day). From each patch, we carefully eliminated magnetic concentrations that were not related to the target ARs.

\begin{figure*}[htbp]
\begin{center}
\includegraphics[width=160mm]{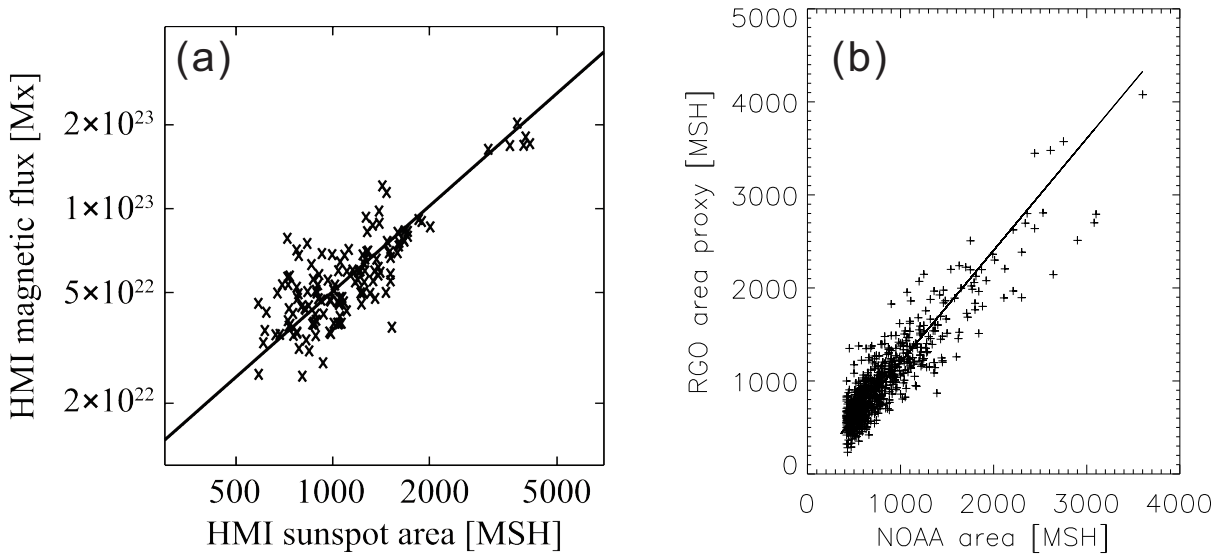}
\end{center}
\caption{
(a) Scatter plot between sunspot areas and total unsigned magnetic fluxes in ARs, both taken from the SHARP data, with the linear fitting to the log-log plot: see Equation (\ref{eq:conversion}).
(b) Comparison of sunspot areas reported by NOAA vs. the database of \citet{man20} that reproduces the RGO area scale. The number of samples is 762. The solid line shows the regression line [Equation (\ref{eq:calibration})]
}
\label{fig:calib}
\end{figure*}

In order to evaluate the total unsigned magnetic flux in maxwell (Mx) from the list of the sunspot areas, we investigated the relation between the sunspot area, $S_{\rm HMI}$, and the total unsigned radial flux, $\Phi_{\rm HMI}$, which can also be calculated from the SHARP data. Note that we measured the total flux only within the smooth bounding curve of the SHARP data to minimize the possibility of including flux that was not related to the target AR and to reduce the noise effect. Figure \ref{fig:calib}(a) shows a scatter plot between $S_{\rm HMI}$ and $\Phi_{\rm HMI}$. From the least-squares fitting to this double logarithmic plot, we obtained the relation between the two parameters,
\begin{eqnarray}
  \log{(\Phi_{\rm HMI}\ {\rm [Mx]})}
  &=&(1.010\pm 0.132)
  \times\log{(S_{\rm HMI}\ {\rm [MSH]})}
  \nonumber\\
  &+&(19.676 \pm 0.400),
  \label{eq:conversion}
\end{eqnarray}
($\log$ means $\log_{10}$; for natural logarithm we use ``$\ln$'' in this paper) where we assumed that both $\Phi_{\rm HMI}$ and $S_{\rm HMI}$ have errors \citep{dem43,pre92}. The obtained conversion equation shows that the total flux is almost linearly related to the sunspot area, $\Phi /S \simeq$ 1660 gauss (AR flux/sunspot area for $S \simeq$ 500 MSH). For comparison, \citet{sch94} derived AR flux/AR area $\simeq$ 150 gauss.

\subsection{Sunspot Area Calibration}
\label{sec:area_calib}
It is known that the USAF/NOAA sunspot areas, $S_{\rm NOAA}$, are systematically smaller than the RGO ones, $S_{\rm RGO}$. In general, the multiplication of the USAF/NOAA values by 1.4--1.5 gives better agreement with the RGO values \citep[e.g.,][]{fli97,hat02,bal09,hat15}. However, \citet{fou14} showed that the inconsistency between the RGO and USAF/NOAA data sets was mainly due to very small sunspots ($\lesssim 2$ MSH) and that the areas for $>300$ MSH equalized. In order to examine if the NOAA records of larger ARs show systematically smaller values, we used the database developed by \citet{man20} which was calibrated to give the same area scale as RGO and was extended to 2021\footnote{http://www2.mps.mpg.de/projects/sun-climate/data/indivi\_group\_area\_1874\_2021\_Mandal.txt}. Among our 820 NOAA regions we excluded 58 regions which we suspect that NOAA and \citet{man20} used different group definitions.

Figure \ref{fig:calib}(b) shows the comparison of 762 regions. We found that mean and standard deviation of the area ratios, $S_{\rm RGO}/S_{\rm NOAA}$, are 1.20 $\pm$ 0.01. Therefore, we simply assumed that the RGO data sets provide reliable values and the NOAA values are converted by
\begin{equation}
  S=S_{\rm HMI}=S_{\rm RGO}= 1.20 \times S_{\rm NOAA}.
  \label{eq:calibration}
\end{equation}
\citet{man20} obtained the conversion factor of 1.48, but the method of comparison is different; they compared daily data while our data are region-wise, maximum-development areas. We have also made the analysis adopting a conversion factor of about 1.4 and found basically the same results, although detailed numerical values changed.

Under the assumption of Equation (\ref{eq:calibration}), we then applied Equation (\ref{eq:conversion}) to the maximum-development sunspot areas $S_{\rm md}$ of both RGO and USAF/NOAA to generate the database on the maximum-development magnetic flux $\Phi_{\rm md}$. A 500 MSH sunspot corresponds to a magnetic flux of $2.53 \times 10^{22}$ Mx. The largest sunspot so far, RGO 14886 in 1947 April (6132 MSH), is estimated to have a total flux of $3.18 \times 10^{23}\ {\rm Mx}$. Flux estimations of modern events such as NOAA 9169, 9393, 10486, and 12192 showed good agreements with the independent measurements by, e.g., \citet{tia08}, \citet{smy10}, \citet{cri09}, \citet{zha10}, and \citet{sun15}.

\begin{figure*}[htbp]
\begin{center}
\includegraphics[width=80mm]{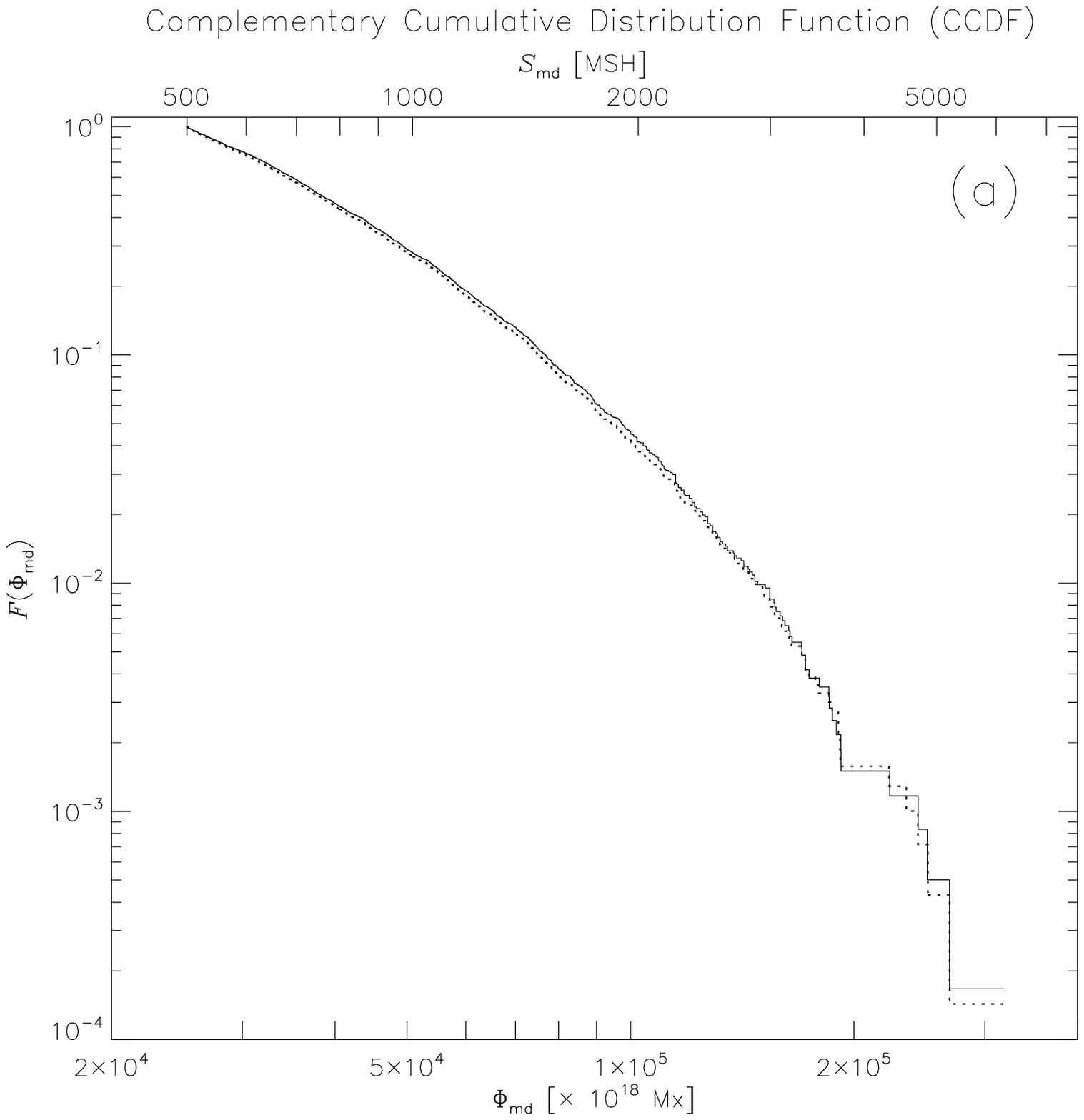}
\includegraphics[width=80mm]{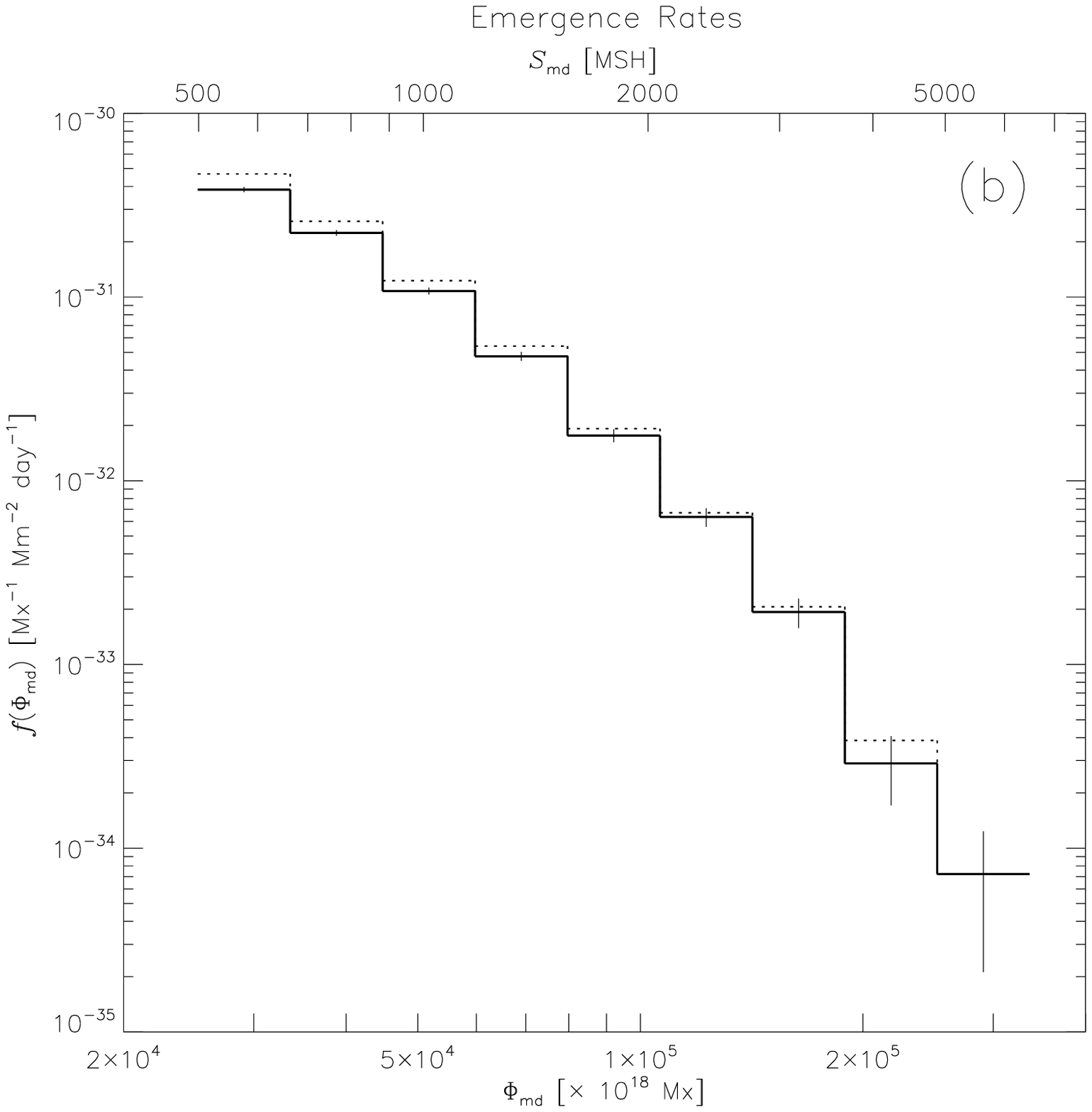}
\end{center}
\caption{
(a) The complimentary cumulative distribution function (CCDF) $F(\Phi)$ of magnetic flux $\Phi_{\rm md}$ [Mx] of large sunspot regions ($S_{\rm md} \geq 500\ {\rm MSH}$) at their maximum development, observed from 1874 April to the end of 2020 (146.7 years). The top horizontal axis shows the corresponding sunspot areas. The dotted curve represents the distribution when recurrent regions not at their maximum development are all counted.
(b) A histogram representing the distribution function $f(\Phi_{\rm md})$ of emergence rates of magnetic flux $\Phi_{\rm md}$ of large sunspot regions, per bin width, unit area in Mm$^2$, and day. The bin width is $\Delta\log \Phi_{\rm md} = 0.125$. Vertical bars indicate the $\sqrt{N}$ uncertainty of $N$ regions contained in each bin (the last bin only contains two regions). The dotted histogram represents the distribution when recurrent regions not at their maximum development are all counted.
\label{fig:data}
}
\end{figure*}

\section{Statistical Analysis}
\label{sec:fitting_models}
\subsection{Definitions}
Our data are made of $N_{\rm s}$ = 2995 values of sunspot magnetic flux at their maximum development, $\Phi_{{\rm md},i}$ ($i=1,2, \ldots , N_{\rm s}$) (subscript ``s'' is meant for ``sources'' or ``source flux tubes''). In the following we simply designate $\Phi_{\rm md}$ as $\Phi$ unless we have to distinguish between maximum-development values and other cases. The minimum value of $\Phi_i$'s is $\Phi_1 \def \Phi_{\rm min} = 2.53 \times 10^{22}$ Mx, corresponding to an area of 500 MSH. As a histogram representation of data depends on how one defines the data bins \citep{cla09}, we will work on the complementary cumulative distribution function [CCDF, represented by $\CCDF (\Phi)$] shown in Figure \ref{fig:data}a, which is uniquely defined in terms of a given observational data set [Equation (\ref{eq:CCDF_obs})]. CCDF is a decreasing function of its argument while the cumulative distribution function (CDF) is an increasing function; therefore, the usage of CCDF is intuitively more straightforward in comparing with the probability distribution function that is also a decreasing function of its argument in the present case.

The slope (or derivative) of CCDF is the usual probability distribution function [PDF, represented by $\PDF (\Phi)$], but here we introduce dimensional parameters and define the flux emergence rate $f (\Phi)$ in units of regions~Mx$^{-1} {\rm Mm}^{-2} {\rm d}^{-1}$ as (Figure \ref{fig:data}b)
\begin{equation}
f (\Phi ) = \PDF (\Phi) \frac{N_{\rm s}}{A T}, \quad \PDF (\Phi) = -\frac{{\rm d} \CCDF(\Phi)}{{\rm d} \Phi},
\label{eq:emergence_rate}
\end{equation}
where $A$ stands for the area of observation (the full Sun, e.g. $6.2 \times 10^{6}\ {\rm Mm}^{2}$ in Sections \ref{sec:fitting_results} and \ref{sec:comparison} and the hemisphere in Sections \ref{sec:instantaneous}) and $T$ =146.7 years $= 5.36 \times 10^4$ days. A histogram is generated showing $f (\Phi_j) = N_j/(A T \Delta\Phi_j)$, where $j$ stands for the bin number, and $N_j$ is the number of sunspot groups with flux values ranging from $\Phi_j$ to $\Phi_j+\Delta\Phi_j$. The flux bins are taken equi-distant in $\log \Phi$ ($\Delta\log \Phi =$ constant = 0.125), so that $\Delta\Phi_j = \Phi_j \ln (10) \Delta (\log \Phi_j)$. The quantity $N_{\rm s} = \Sigma_j N_j$ is explicitly used when we compare the observed histogram with a theoretical distribution function $P$.

%
In Figures \ref{fig:data}a and \ref{fig:data}b, the solid curves are from the maximum development areas while the dotted curves show the cases when we did not exclude the recurrent regions other than their maximum area development. We can expect that a larger region may appear multiple times at smaller areas and increase the counts at lower bins. The effect of not excluding recurrent regions is to slightly steepen the slope of the distribution.

\begin{figure*}
\begin{center}
\includegraphics[width=80mm]{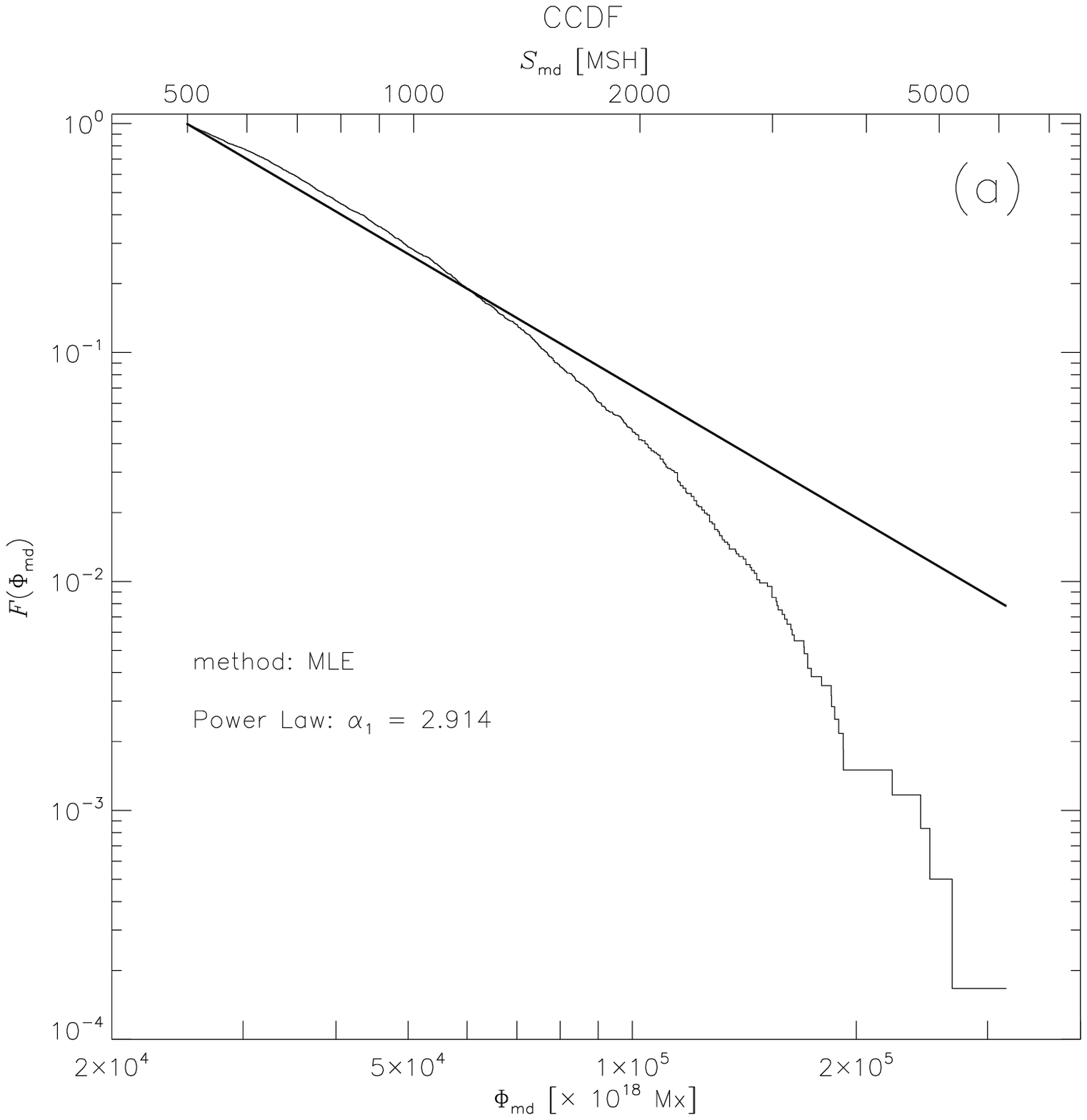}
\includegraphics[width=80mm]{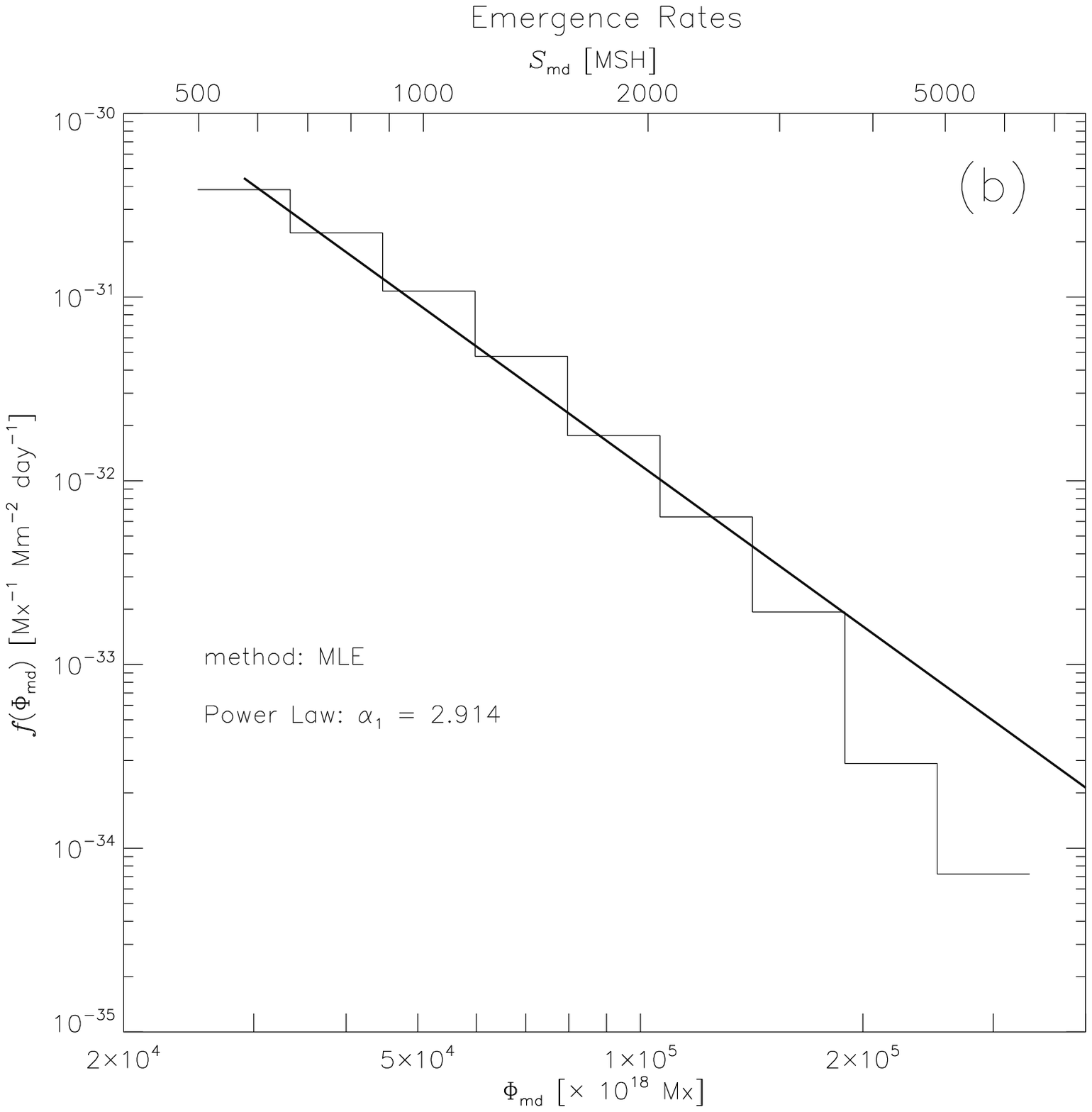}
\end{center}
\caption{
(a) An MLE power-law fit (straight line) to the complimentary cumulative distribution function of magnetic flux $\CCDF (\Phi)$.
(b) The power-law fit of panel (a) for $\CCDF (\Phi)$ is converted to the probability distribution function $f(\Phi)$ and is overplotted on the histogram of the data.
\label{fig:power-law}
}
\end{figure*}

\begin{table}[htbp]
\begin{center}
\caption{
Derived parameters and 1-$\sigma$ error ranges
}
\label{tab:summary}
\begin{tabular}{llllrrrr}
\hline
\hline
Model & Method & \multicolumn{2}{c}{Parameter values and errors} & $\Delta$AIC$^a$ & $\sqrt{N_{\rm s}}$ KS$^b$
 & \multicolumn{1}{l}{KS} & \multicolumn{1}{l}{KSr} \cr
      &        &    &                                            &                 &           
 & \multicolumn{1}{l}{prob.$^c$} & \multicolumn{1}{l}{P-value$^d$} \cr
\hline
\multicolumn{2}{l}{Power law} \cr
& MLE     & $\alpha_1 = 2.914 \pm 0.035$ &                            & 184.6 & 4.14 & $< 10^{-6}$ & 0.019 \cr
\multicolumn{2}{l}{Tapered power law} \cr
& MLE     & $\alpha_2 = 1.808 \pm 0.091$ & $\beta_2 = 0.669 \pm 0.056$ & 2.79 & 1.059 & 0.210 & 0.290 \cr
& KSr=min & $\alpha_2 = 1.886 \pm 0.153$ & $\beta_2 = 0.614 \pm 0.083$ & 3.77 & 1.211 & 0.105 & 0.444 \cr
\multicolumn{2}{l}{Gamma} \cr
& MLE     & $\alpha_3 = 1.358 \pm 0.138$ & $\beta_3 = 0.591 \pm 0.057$ & 1.51 & 0.989 & 0.279 & 0.458 \cr
& KSr=min & $\alpha_3 = 1.450 \pm 0.224$ & $\beta_3 = 0.547 \pm 0.083$ & 2.11 & 1.08 & 0.195 & 0.665 \cr
\multicolumn{2}{l}{Lognormal} \cr
& MLE     & $\mu     = -0.300 \pm 0.099$ & $\sigma  = 0.777 \pm 0.035$ & 1.55 & 0.630 & 0.817 & 0.414 \cr
& KSr=min & $\mu     = -0.157 \pm 0.188$ & $\sigma  = 0.728 \pm 0.061$ & 4.27 & 1.075 & 0.196 & 0.490 \cr
\multicolumn{2}{l}{Weibull} \cr
& MLE     & $k =        0.625 \pm 0.044$ & $\beta_5 = 2.278 \pm 0.217$ & 0.00 & 0.827 & 0.497 & 0.911 \cr
& KSr=min & $k =        0.629 \pm 0.068$ & $\beta_5 = 2.270 \pm 0.396$ & 0.06 & 0.868 & 0.434 & 0.850 \cr
\hline
\hline
\end{tabular}
\end{center}
$^a$: Relative AIC values with respect to the minimum value of all the AIC values.\\
$^b$: Kolmogorov-Smirnov statistic KS derived from the observed data and the assumed model, multiplied by $\sqrt{N_{\rm s}}$. \\
$^c$: Theoretical probability that the KS metric shows values larger than the observed KS metric. \\
$^d$: Probability based on the simulation runs that the simulated KSr values are larger than the observed KSr value.
\end{table}

\subsection{Models and Fitting Procedures}
In this paper we will try to fit the observed data set $\Phi_i$ by
\begin{itemize}
\item[(1)] a power-law distribution (parameter: $\alpha_1$),
\item[(2)] a tapered power-law distribution (parameters: $\alpha_2, \beta_2$),
\item[(3)] a truncated gamma distribution (parameters: $\alpha_3, \beta_3$),
\item[(4)] a truncated lognormal distribution (parameters: $\mu, \sigma$), and
\item[(5)] a truncated Weibull distribution (parameters: $k, \beta_5$).
\end{itemize}
A tapered power-law distribution (also called the Pareto distribution of the third kind \citep{jkb94} in contrast to the original Pareto distribution which is a pure power law) has a CCDF that is a product of a power law and an exponential function. A gamma distribution has a PDF that is a product of a power law and an exponential function, and its CCDF is represented in terms of the gamma function. These have been used to describe the distribution functions of earthquake magnitudes \citep[e.g.][]{kag02,ser17}, in comparison with the power-law distribution which is named the Gutenberg-Richter relation in seismology \citep[e.g.][]{uts99}.

The power-law distribution is a one-parameter model while the other four are two-parameter models. Models (1), (2), and (3) have a power-law component with the exponent $\alpha_1$, $\alpha_2$, and $\alpha_3$, respectively. Models (2), (3), and (5) have an exponential component whose decay coefficients are represented by $\beta_2$, $\beta_3$, and $\beta_5$. The best-fit parameter values can be given systematically by the maximum-likelihood estimator (MLE), by maximizing the log-likelihood (LLH),
\begin{equation}
{\rm LLH} = \sum \ln (\PDF (\Phi_i)).
\end{equation}
The definitions of these five distributions and their MLE solutions are given in Appendix A.

Whether one model is more superior to the others can be assessed by comparing the AIC values \citep{aka74},
\begin{equation}
{\rm AIC} = -2 \times \mbox{max(LLH)} + 2 K,
\label{eq:AIC}
\end{equation}
where $K$ is the number of parameters ($K=1$ or 2 in our analysis). (In another often-used criteria, Bayesian information criteria or BIC, $K \ln n$ ($n$ is the sample size) is used instead of $2K$; \cite{bur02}.) By increasing the number of parameters, the fitting becomes better and LLH increases. However, the introduction of more parameters is not justified if AIC does not decrease. If the AIC value of one model is smaller than the AIC of the other by 9--11 or more, the former model is regarded better than the latter \citep{bur11}.

AIC is a relative measure, and it is possible that the model preferred by AIC still gives a poor fit. Whether the fitting by one model is not satisfactory and should be rejected can be estimated by evaluating a statistical measure that quantifies the difference between the observed and modeled CCDFs and by comparing that measure with a theoretical threshold \citep{ste70, ste16}. Here we use the Kolmogorov-Smirnov statistic KS defined by Equation (\ref{eq:KS}). For a large-enough $N_{\rm s}$, the probability for the observed $\sqrt{N_{\rm s}}$ KS value or larger to be obtained is theoretically given as a function (Kolmogorov-Smirnov function) which has relatively weak dependence on $N_{\rm s}$. Alternatively we can utilize simulation runs to estimate the probability (see below).

Once the parameter values are obtained, we can estimate their error ranges using the parametric bootstrap method \citep{efr93,bur02} as follows.
\begin{itemize}
\item[(1)] Generate $N_{\rm s}$ sets of uniform random variables $y_i$, $i=1,2, \ldots , N_{\rm s} \ (0 \le y_i \le 1)$.
\item[(2)] Find $\Phi_i$ such that $\CCDF (\Phi_i) = y_i$.
\item[(3)] For the data set $\Phi_i$ ($i=1,2, \ldots , N_{\rm s}$), obtain the MLE solutions for the parameters and the KS statistic.
\item[(4)] Repeat (1)--(3) $m$ times; then the distributions of the parameter values and the KS statistic are obtained.
\item[(5)] Evaluate the 1-$\sigma$ width of these distributions and adopt them as the error ranges of the parameters.
\item[(6)] For the KS statistic, if the number of cases where the KS values exceed the observed value KS$_{\rm obs}$ is $m'$, then $m'/m$ gives the probability (P-value) that such a value of KS$_{\rm obs}$ is obtained under the assumed model.
\end{itemize}
The 1-$\sigma$ values defined in step (5) scale basically as $1/\sqrt{ N_{\rm s}}$ and do not depend on $m$ if it is taken sufficiently large (we used $m=10^5$).

The procedure described above has some points to consider. First, the MLE solution is not intended to geometrically fit a model CCDF to the observed CCDF. Particularly, it does not care much about the fitting at the tail, because the MLE solution is mostly determined by data points of highest density, i.e. near to the lower end of the distribution. A more geometrically-favorable solution may be obtained by, say, minimizing directly the KS statistic. Second, the KS statistic is not sensitive to misfitting at the lower and upper ends of the distribution, because by definition both the observed and model CCDF match (taking the values of 0 or 1) at the ends. It is known that the contributions to KS roughly scales as $1/\sqrt{\CCDF (1-\CCDF )}$ \citep{and52}, and \citet{cla09} suggested to use a modified form of the KS statistic by dividing its components by $1/\sqrt{\CCDF (1-\CCDF )}$ to enhance the sensitivity of the test at both ends. Here we introduce a KSr (revised KS) statistic defined by Equation (\ref{eq:KSr}), by dividing only by $1/\sqrt{\CCDF}$ to enhance the sensitivity at the tail.

In summary, we will apply two methods.
\begin{itemize}
\item Method 1: Seek the MLE solution, check the AIC values, and test the KS statistic by the Kolmogorov-Smirnov function. Then by simulation runs, estimate the error ranges of the obtained parameters, and the P-value of the observed KSr statistic.
\item Method 2: Minimize KSr to obtain the solution, check the AIC values, and test the KS statistic by the Kolmogorov-Smirnov function. Then by simulation runs, estimate the error ranges of the obtained parameters, and the P-value of the observed KSr statistic.
\end{itemize}
Method 2 was not applied to the power-law model because the deviations of the model at the tail are large and the introduction of KSr may not make sense.

\begin{figure*}[htbp]
\begin{center}
\includegraphics[width=80mm]{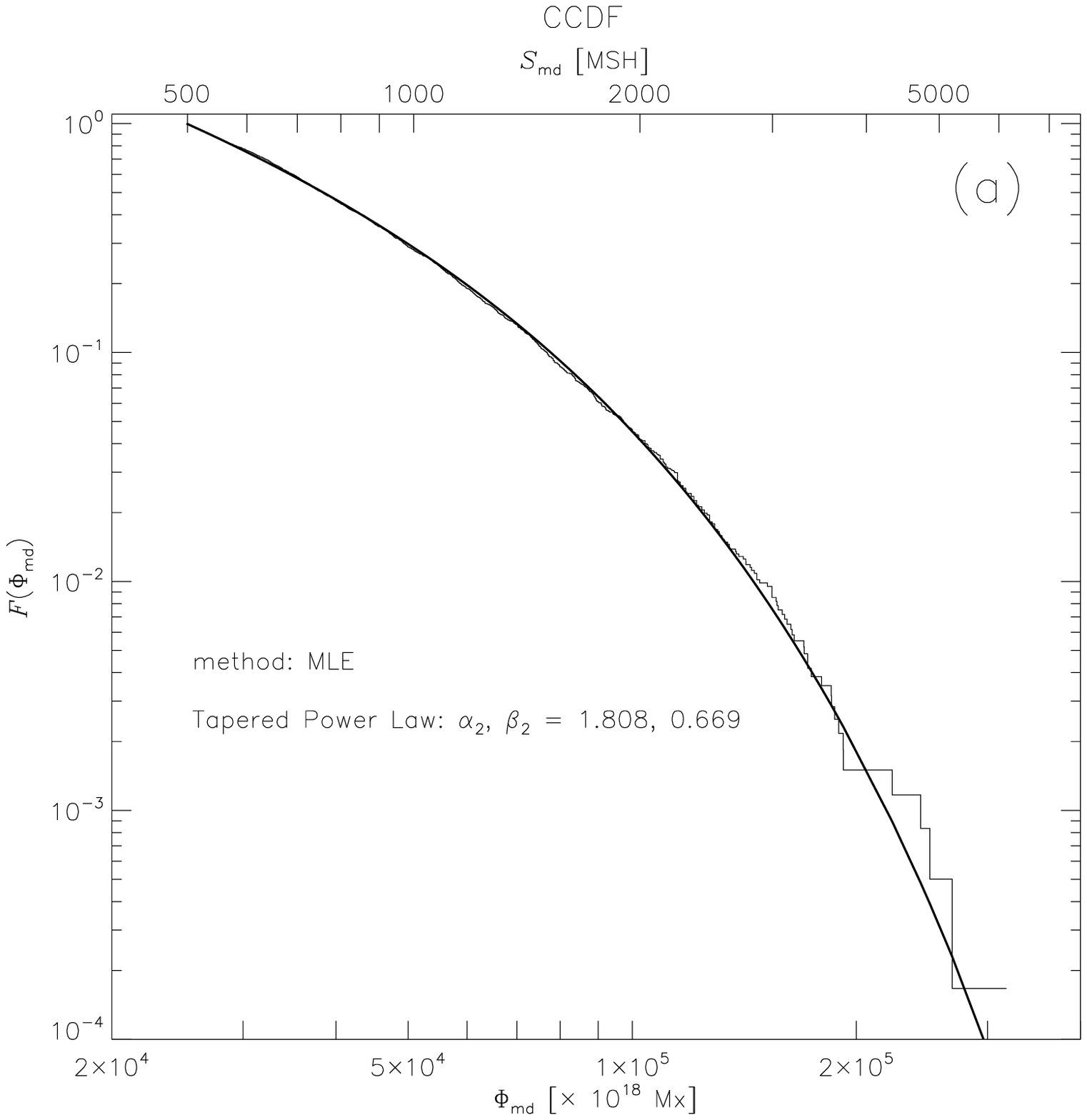}
\includegraphics[width=80mm]{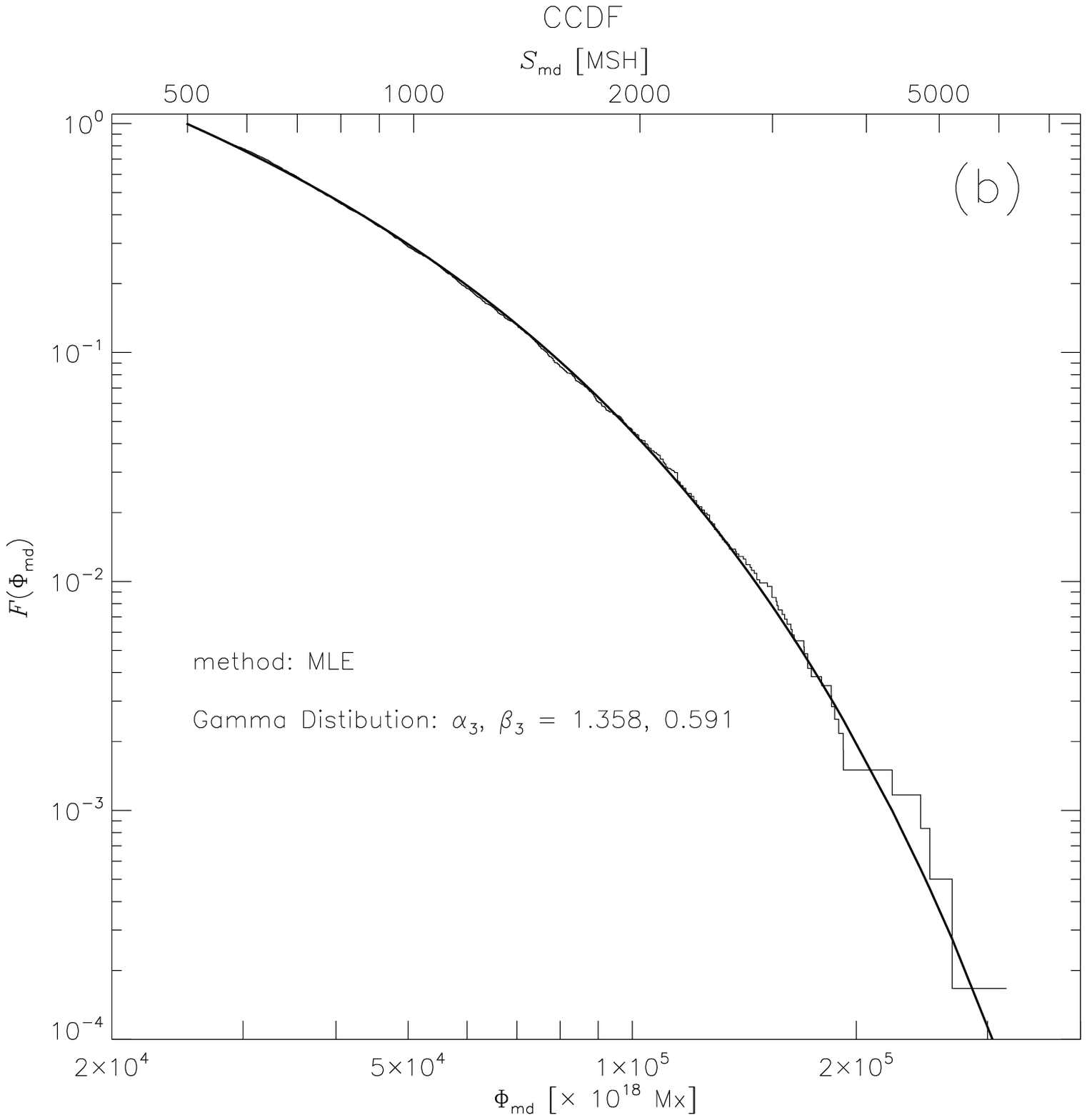}
\includegraphics[width=80mm]{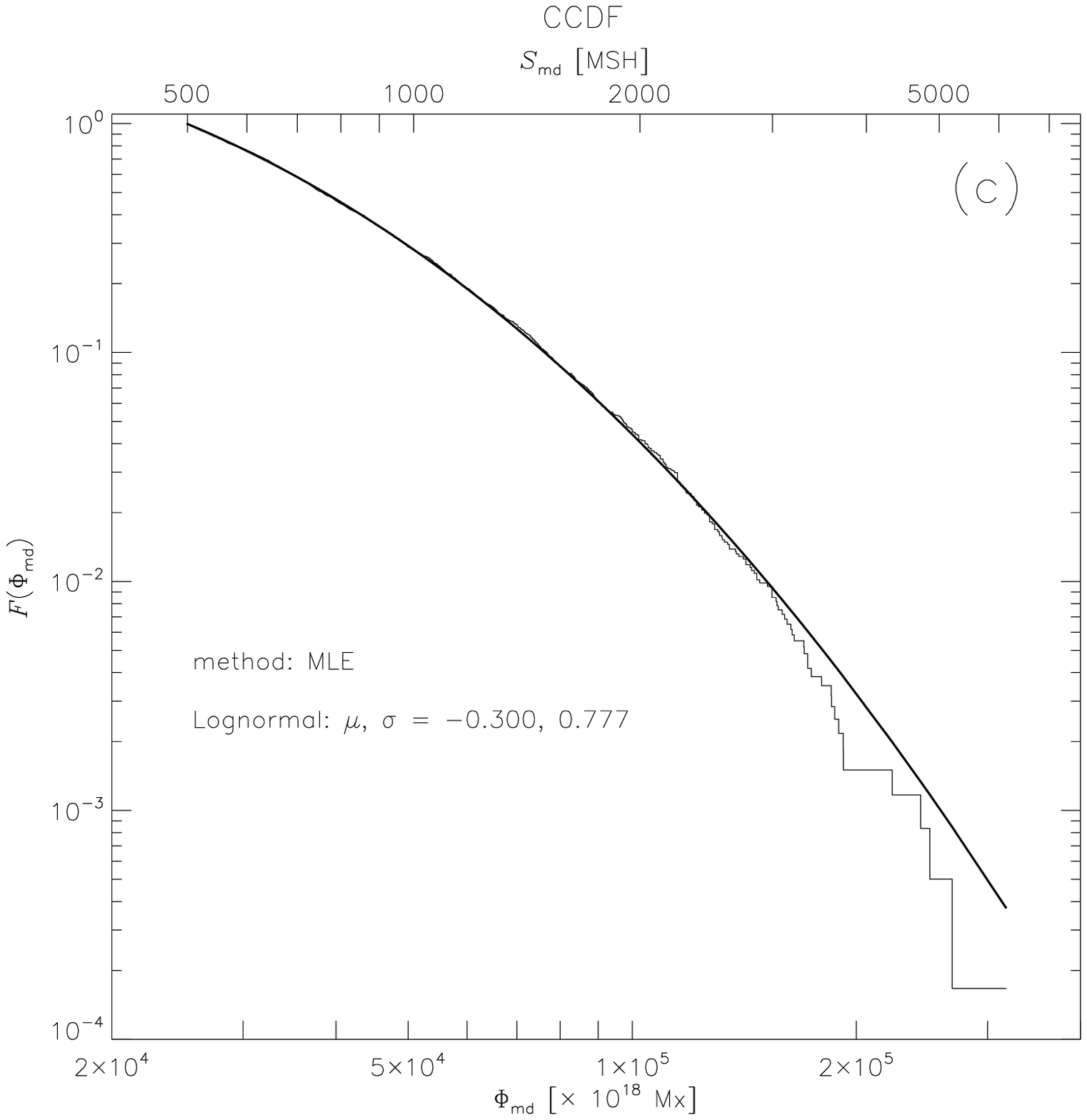}
\includegraphics[width=80mm]{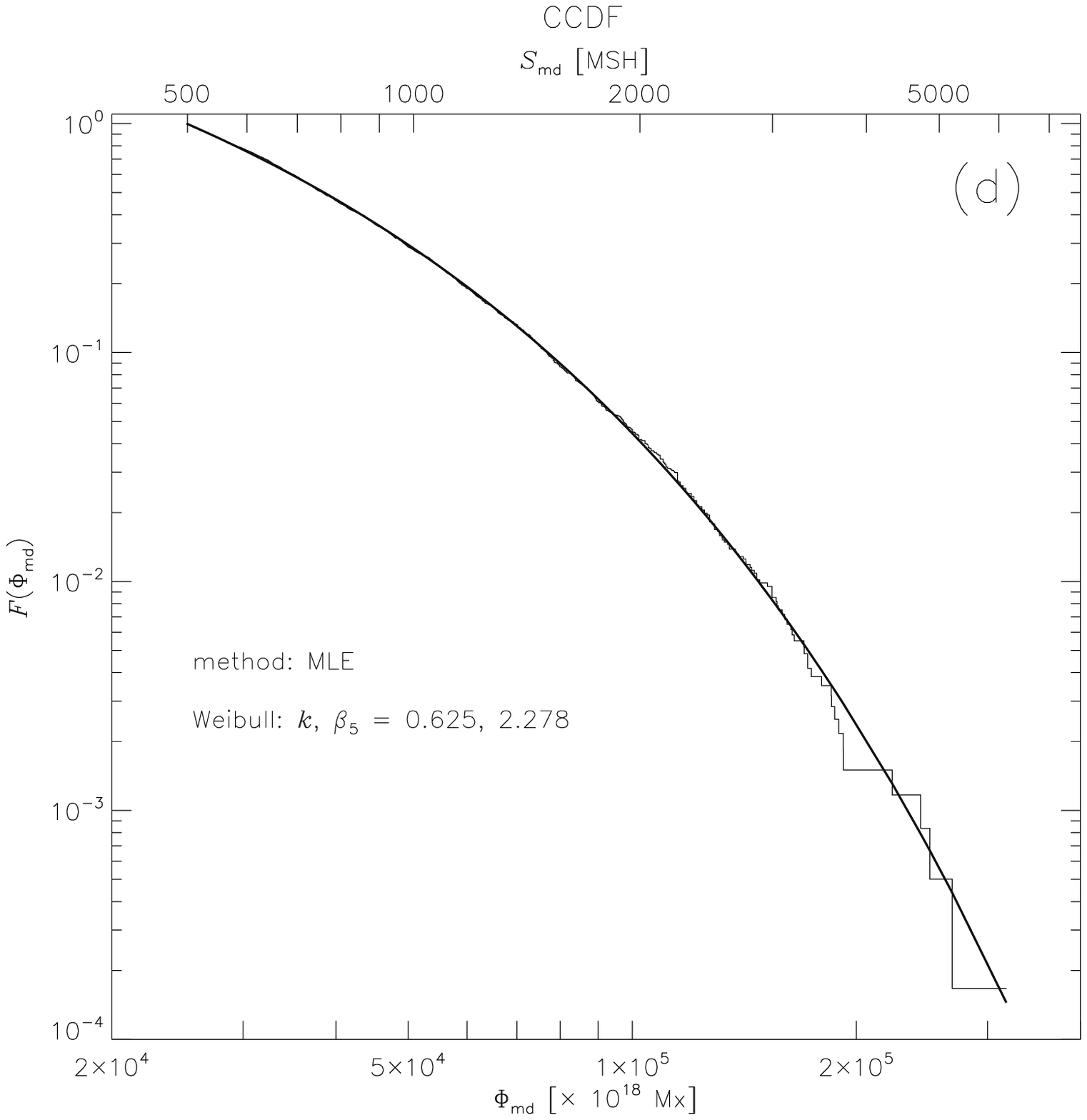}
\end{center}
\caption{
MLE fits to the complimentary cumulative distribution functions $\CCDF (\Phi)$ using models of (a) tapered power law, (b) gamma, (c) lognormal, and (d) Weibull distributions.
\label{fig:all-models}
}
\end{figure*}

\section{Fitting Results}
\label{sec:fitting_results}
\subsection{Power Law}
Figure \ref{fig:power-law} shows the results of power-law fitting. The MLE solution for the exponent is $\alpha_1 =2.91$, and the KS statistic is large, $\sqrt{N_{\rm s}}$ KS = 4.1. Therefore, the probability of obtaining such a value or larger is infinitesimally small, and the model is safely rejected.

\begin{figure*}
\begin{center}
\includegraphics[width=80mm]{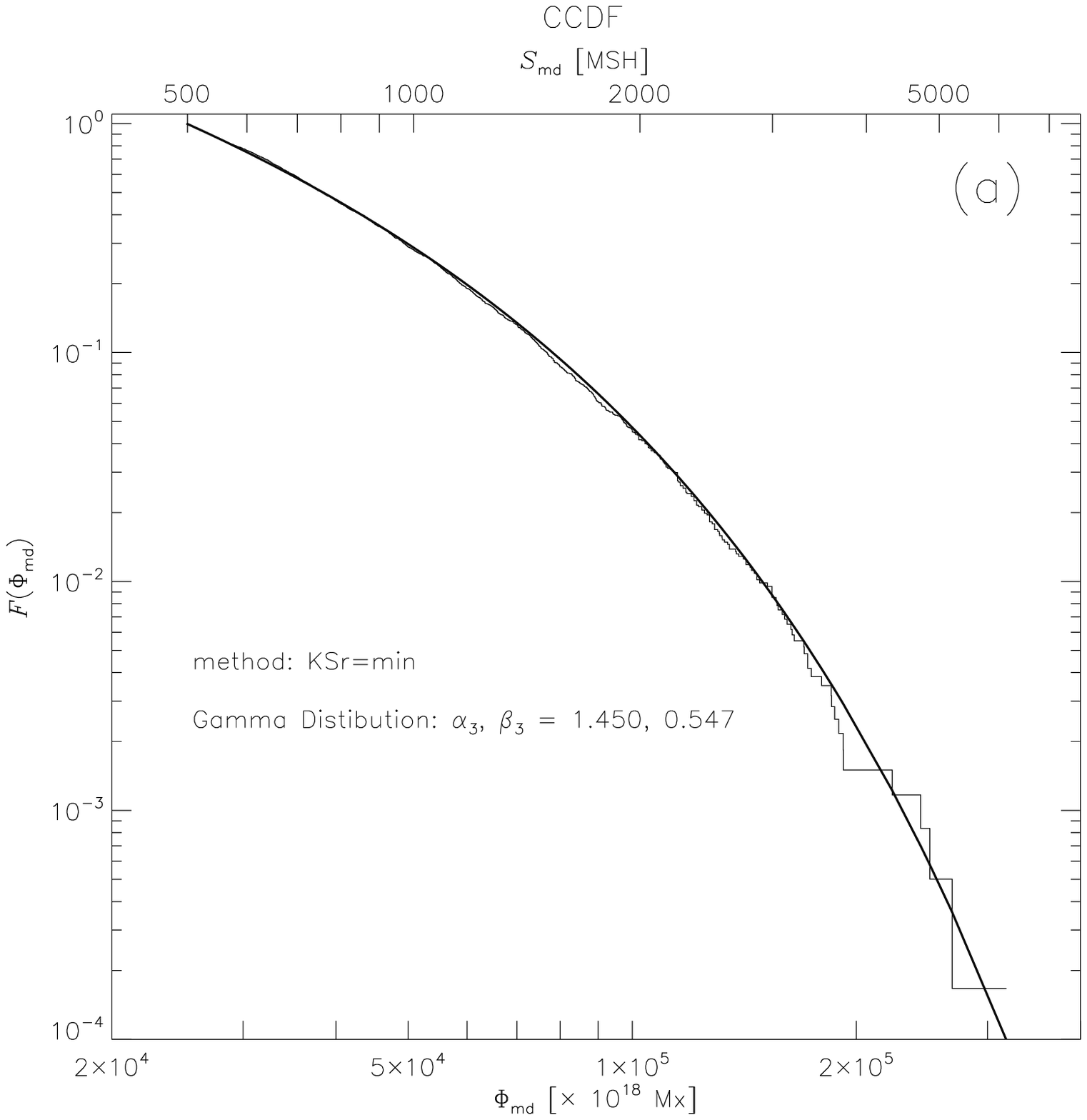}
\includegraphics[width=80mm]{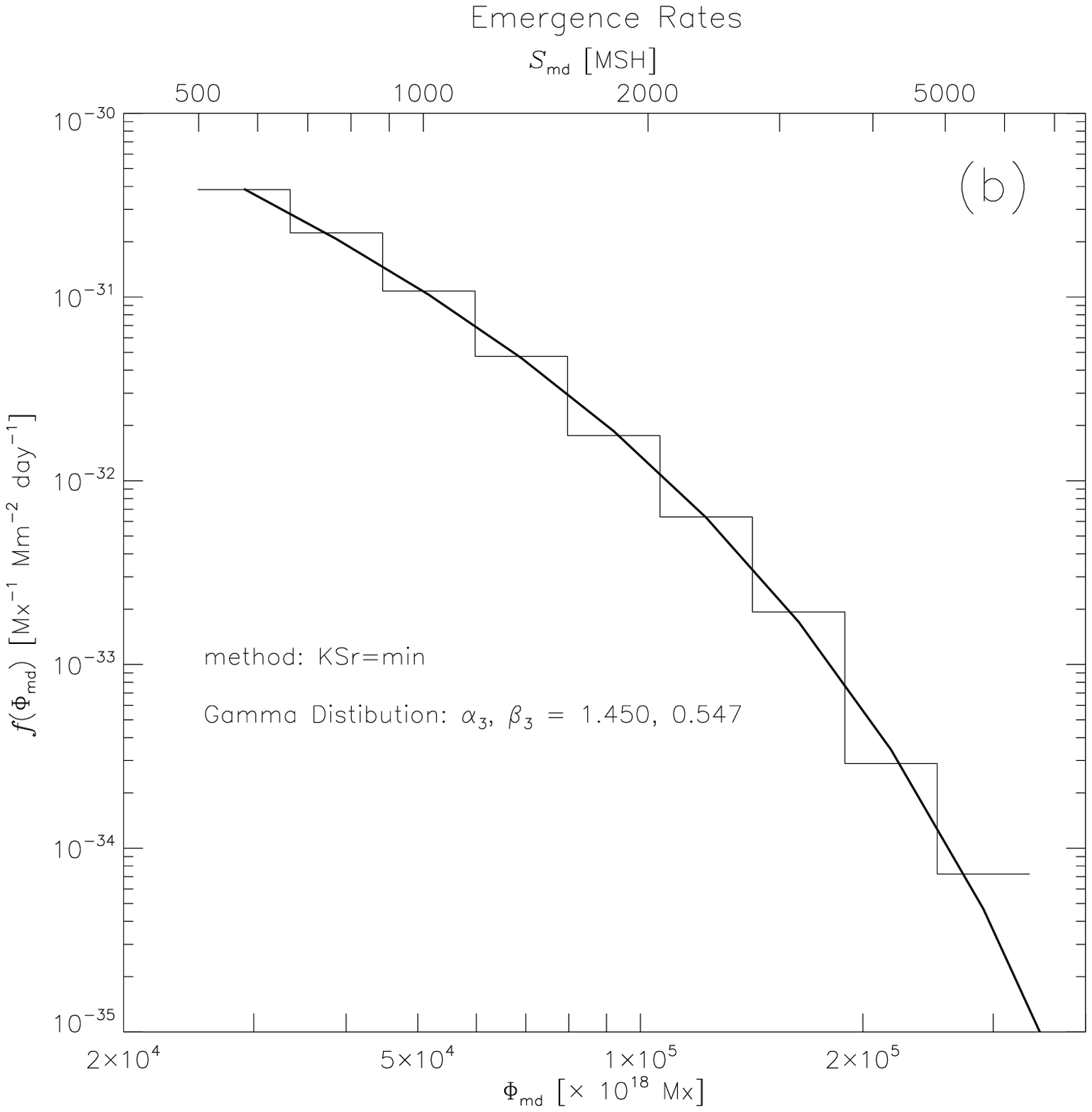}
\end{center}
\caption{
(a) A fit to the complimentary cumulative distribution function (CCDF) $F(\Phi_{\rm md})$ using the gamma distribution model and by minimizing the KSr metric.
(b) The gamma-distribution fit of panel (a) for $\CCDF (\Phi_{\rm md})$ is converted to the probability distribution function $f(\Phi_{\rm md})$ and overplotted on the histogram of the data.
\label{fig:gamma-function}
}
\end{figure*}

\subsection{Two-Parameter Models}
Figures \ref{fig:all-models} (a)--(d) show the results of fitting by adopting MLE solutions (Method 1). Method 2 also gives similar plots. In Figure \ref{fig:gamma-function}, both the CCDF and emergence rates $f(\Phi_{\rm md})$ together with the fitted functions are given by applying Method 2 to the gamma distribution model, which is our most favorable model.

Table \ref{tab:summary} summarizes, for MLE and KSr=min (except for power law) methods respectively, the obtained parameter values (with error ranges), relative AIC values, KS statistics and its theoretical probability, and P-values based on the KSr statistic. Here $\Delta$AIC ($\Delta$BIC = $\Delta$AIC among the two-parameter models) means the values of AIC with respect to its smallest value in the models (which happened to take place for the MLE model applied to the Weibull distribution). Strictly speaking, AIC is defined when LLH is maximized [the MLE solution, Equation (\ref{eq:AIC})], but we also applied the same formula by replacing max(LLH) with LLH from a particular solution not maximizing LLH. Therefore, for each model, $\Delta$AIC is smaller for the MLE solution than for the solution with KSr=min. Likewise, the P-value based on KSr is larger (better fitting) for the solution with KSr=min compared to the MLE solution.

\begin{figure*}[htbp]
\begin{center}
\includegraphics[width=180mm]{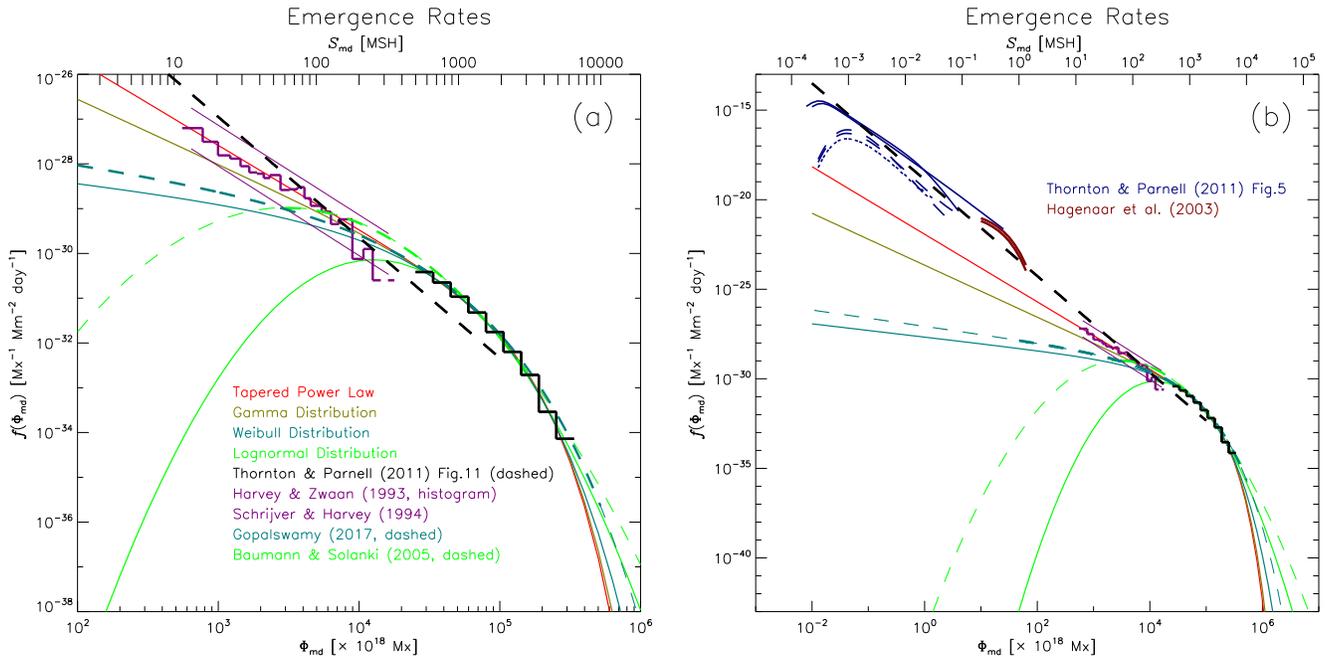}
\end{center}
\caption{
(a) Comparison of magnetic flux emergence rates $f(\Phi_{\rm md})$. The histogram drawn with a thick black line is our data. Four kinds of fits by minimizing the KSr metric are shown, respectively, for tapered power-law (red), gamma (olive), Weibull (teal green), and lognormal (lime green) distributions. The black dashed line of power-law exponent 2.69 is from \citet{tho11}. The histogram shown in purple is from \citet{khz93}, and the two parallel lines in purple are the power laws of exponent=2 \citep{sch94} representing their amplitude ranges in solar activity minima/maxima. The dashed curve in lime green is the lognormal fit from \citet{bau05}. The dashed curve in teal green is the Weibull function fit from \citet{gop18}.
(b) Same as panel (a) but the plotting range is extended to smaller flux values. Newly added graphs are the SOHO/MDI data from \citet[][maroon]{hst03} and the Hinode/SOT data from \citet[][navy blue]{tho11}.
\label{fig:comparison}
}
\end{figure*}

The following properties can be found on this table.
\begin{enumerate}
\item The AIC values of the two-parameter models are much smaller than the case of the power law, so that all these four models are better than the power law. The four models show $\Delta$AIC values less than 5 and cannot be discriminated.
\item From the KS probabilities, the MLE solutions show better performance but the KSr=min solutions are also acceptable.
\item From the P-values of the KSr metric, the solutions minimizing KSr show better performance but the MLE solutions are also acceptable.
\item The power-law indices of the tapered power law and the gamma distributions are $\alpha_2 =$ 1.8--1.9 and $\alpha_3 =$ 1.35--1.45. The reason why $\alpha_2>\alpha_3$ is given in Appendix \ref{appendix:tapered-power-law}; the tapered power-law distribution actually contains a mixture of two exponents $\alpha_2$ and $\alpha_2-1$, and its overall behavior is somewhere between them. If we extend the PDF toward smaller $\Phi_{\rm md}$ values, the tapered power-law distribution asymptotically approaches the power law with exponent $\alpha_2$. In any case it is important to point out that both distributions show the power-law-like behavior with exponent less than 2, namely the overall contributions to the magnetic flux supply come mostly from large $\Phi_{\rm md}$ regions.
\item As we discuss later (Figure \ref{fig:comparison}), the behavior of lognormal and Weibull distributions extended to smaller values of $\Phi_{\rm md}$ is different from the tapered power-law and gamma distributions (the latter two behave essentially like a power law with exponent $\simeq$ 1.35--1.9). The Weibull distributions approach toward a very flat power-law distribution $\propto \Phi_{\rm md}^{-0.4}$, and the lognormal distributions decrease toward small $\Phi_{\rm md}$. Therefore, our preferred models are the tapered power-law and gamma distributions.
\item The tapered power-law and gamma distributions show a steep fall-off for large $\Phi_{\rm md}$ values, steeper than power laws, indicating that the probability of having extremely large active regions is vanishingly small and there is a practical upper limit in the size and magnetic flux of emerging active regions.
\end{enumerate}

\section{Comparison with Published Results}
\label{sec:comparison}
In this section, we will compare our results on the two-parameter models with the published observational data and fitting results. In Figure \ref{fig:comparison}a which shows the flux emergence rates $f(\Phi_{\rm md})$, the histogram in a thick black line is our data (500 MSH $\leq S_{\rm md} \leq$ 6132 MSH, $2.53 \times 10^{22}\ {\rm Mx} \leq \Phi_{\rm md} \leq 3.18 \times 10^{23}\ {\rm Mx}$). Four solid curves show our fits by minimizing the KSr metric; tapered power-law (red), gamma (olive), Weibull (teal green), and lognormal (lime green) distributions. They are extended down to $\Phi_{\rm md} = 10^{20}$ Mx ($S_{\rm md} \simeq 2$ MSH) and up to $\Phi_{\rm md} = 10^{24}$ Mx ($S_{\rm md} \simeq 19000$ MSH). Toward the smaller ends of $\Phi_{\rm md}$, the tapered power-law, gamma, and Weibull distributions approach power laws with exponents 1.89, 1.45, and 0.37, respectively

The histogram in a thick purple line is taken from \citet{khz93}, who derived the emergence rates of bipolar active regions using the data obtained at NSO Kitt Peak \citep{liv76}. \citet{sch94} reported that this distribution is fitted by a power law with exponent 2, and the amplitudes vary by roughly a factor of 10 between activity minimum and maximum, as shown by two parallel lines in purple.

The thick dashed curve in lime green reproduces the lognormal fit to the RGO data ($S_{\rm md} \geq$ 60 MSH) by \citet{bau05}, extended down to $S_{\rm md} \simeq 2$ MSH (thin dashed curve). Below the peak at $S_{\rm md} \simeq 62.2$, the curve goes down to $f \rightarrow 0$. Our lognormal fit peaks at $S_{\rm md} \simeq$ 250 MSH and then decreases to $f \rightarrow 0$. At least for our lognormal fitting, this decrease is not due to small fluxtubes losing their darkness, because even the smallest regions (500 MSH) in our sample are fairly large regions. Rather, this is a result of the shape of the observed distribution that bends down toward large $\Phi_{\rm md}$, and the derived $\mu$ value may not represent any physical significance. \citet{bog88} gave the values 0.34--0.62 MSH for the peak of the instantaneous distribution function of sunspot umbral areas modeled by lognormal distribution.

The thick dashed curve in teal green reproduces the Weibull fit to the RGO and USAF/NOAA data by \citet{gop18}, which covered all the data $S_{\rm md} \geq 1$ MSH. The curve is very close to our Weibull fit.

In Figure \ref{fig:comparison}b, we extended the plot range to $10^{15}$ Mx $\leq \Phi_{\rm md} \leq 10^{25}$ Mx ($2.3 \times 10^{-5}$ MSH $\lesssim S_{\rm md} \lesssim 1.9 \times 10^{5}$ MSH), and added two more data sets. The thick solid curve in brown is from \citet{hst03}, who investigated the emergence rates of small-scale bipolar magnetic patches (ephemeral regions) using the data from SOHO/MDI. The thick curves in navy blue (solid, dashed, dotted) are from \citet[][Figure 5]{tho11}, who analyzed the emergence rates of small-scale magnetic patches using the data from Hinode/SOT. The thin dashed line in teal green is the downward extension of \citet{gop18}'s Weibull distribution.

The thick dashed lines in black in Figures \ref{fig:comparison}a and \ref{fig:comparison}b show the power law of exponent 2.69 suggested by \citet{tho11} to cover all the way from small-scale flux concentrations to large active regions. Our picture is different from theirs; the flux emergence rates of active regions are characterized by a power-law-like behavior of exponents between 1.45 (olive) and 1.89 (red) in Figure \ref{fig:comparison}. Our results are roughly consistent with the observations by \citet{khz93}, who also showed that the distribution amplitudes varied by a factor of 10 between activity minima and maxima. Ephemeral regions and much smaller flux concentrations may have a power-law distribution with exponent 2.69, but they show little changes \citep[or even anti-phase changes;][]{hst03} with the solar cycle, and they may give way to the active region component somewhere at around $\Phi_{\rm md} \simeq 10^{20}$ Mx. The exponent larger than 2 in small-scale flux concentrations means dominant contributions of flux emergence in the smallest end of the distribution function. Those small-scale flux emergence may be sustained by a local dynamo \citep{cat99,bue13}. The total (unsigned) magnetic flux on the Sun in scales exceeding a few arcseconds ($\gtrsim 1000$ km) varies in phase with the solar cycle \citep{arg02}, and the flux at activity maximum is about four times the flux at minimum. Therefore, flux emergence in small scales would not accumulate to systematically overwhelm the magnetic flux from active regions.

\begin{table}[htbp]
\begin{center}
\caption{
Sunspot area, magnetic flux, and interval of appearance
}
\label{tab:prediction}
\begin{tabular}{rrr}
\hline
\hline
$S$ [MSH] & $\Phi$ [Mx] & Interval [year] \cr
\hline
  500 & $2.5 \times 10^{22}$ & $5.242 \pm 0.001  \times 10^{-2}$ \cr
 1000 & $5.1 \times 10^{22}$ & $1.82 \pm 0.04  \times 10^{-1}$ \cr
 2000 & $1.0 \times 10^{23}$ & $1.20 \pm 0.04  \times 10^{0\phantom{-}}$ \cr
 3000 & $1.5 \times 10^{23}$ & $6.0^{+0.9}_{-0.8}  \times 10^{0\phantom{-}}$ \cr
 6132 & $3.2 \times 10^{23}$ & $5.2^{+3.7}_{-2.2} \times 10^{2\phantom{-}}$ \cr
10000 & $5.2 \times 10^{23}$ & $8.2^{+17}_{-5.5} \times 10^{4\phantom{-}}$ \cr
\hline
\hline
\end{tabular}
\begin{tabular}{rrr}
\hline
\hline
$S$ [MSH] & $\Phi$ [Mx] & Interval [year] \cr
\hline
$6.2^{+0.5}_{-0.4} \times 10^3$ & $(3.2^{+0.2}_{-0.2}) \times 10^{23}$ & $1.0 \times 10^3$ \cr
$7.9^{+0.7}_{-0.5} \times 10^3$ & $(4.1^{+0.4}_{-0.3}) \times 10^{23}$ & $1.0 \times 10^4$ \cr
$9.7^{+1.0}_{-0.7} \times 10^3$ & $(5.0^{+0.5}_{-0.4}) \times 10^{23}$ & $1.0 \times 10^5$ \cr
$1.15^{+0.12}_{-0.09} \times 10^4$ & $(6.0^{+0.6}_{-0.5}) \times 10^{23}$ & $1.0 \times 10^6$ \cr
$1.51^{+0.18}_{-0.14} \times 10^4$ & $(7.9^{+1.0}_{-0.7}) \times 10^{23}$ & $1.0 \times 10^8$ \cr
$1.83^{+0.23}_{-0.17} \times 10^4$ & $(9.6^{+1.2}_{-0.9}) \times 10^{23}$ & $4.6 \times 10^9$ \cr
\hline
\hline
\end{tabular}
\par\smallskip
\parbox[l]{15cm}{%
\small
The ranges of values are based on the 1-$\sigma$ error ranges given in Table \ref{tab:summary}, namely $\alpha_3 = 1.450 \pm 0.224$, $\beta_3 = 0.547 \pm 0.083$.
}
\end{center}
\end{table}

\subsection{Emergence Frequency of Large Sunspots}
\label{sec:expected_life}
Based on the fitting results, we are able to predict the expected frequencies of large sunspots. Table \ref{tab:prediction} shows, by using the model of gamma distribution with parameter values determined from KSr=min ($\alpha_3 =$ 1.450 $\pm$ 0.224, $\beta_3 =$ 0.547 $\pm$ 0.083), the expected appearance intervals as a function of sunspot area (left half of the table), and the expected sunspot areas for the specified values of appearance intervals (right half of the table). The ranges of values given are based on the 1-$\sigma$ error ranges given in Table \ref{tab:summary}. It turned out that the errors in $\alpha_3$ and $\beta_3$ are roughly inversely correlated, so that the ranges of values shown in Table \ref{tab:prediction} correspond to $(\alpha_3, \beta_3)=$ (1.450 + 0.224, 0.547 $-$ 0.083) and (1.450 $-$ 0.224, 0.547 + 0.083).

The time interval for regions with flux larger than $\Phi$ to appear is given by $[\CCDF(\Phi) N_{\rm s}/T]^{-1}$ ($N_{\rm s}$ = 2995, $T=$ 146.7 years). The sunspots larger than 500 MSH are expected to appear at a rate of every 19 days. The regions as large as or exceeding the largest region in our database (6132 MSH) are expected to appear every 520 years. Beyond roughly 10000 MSH, the interval becomes longer than $8 \times 10^4$ years, and even after the lifetime of the Solar System ($4.6 \times 10^9$ years), one can only expect a region as large as 1.8 $\times 10^4$ MSH. These are due to the exponential decline of the probability distribution function. By taking into account the 1-$\sigma$ errors, the expected interval for 10000 MSH regions is reduced to 2.7 $\times 10^4$ years, and 2.1 $\times 10^4$ MSH is the possible maximum size after $4.6 \times 10^9$ years.

\begin{figure*}[htbp]
\begin{center}
\includegraphics[width=140mm]{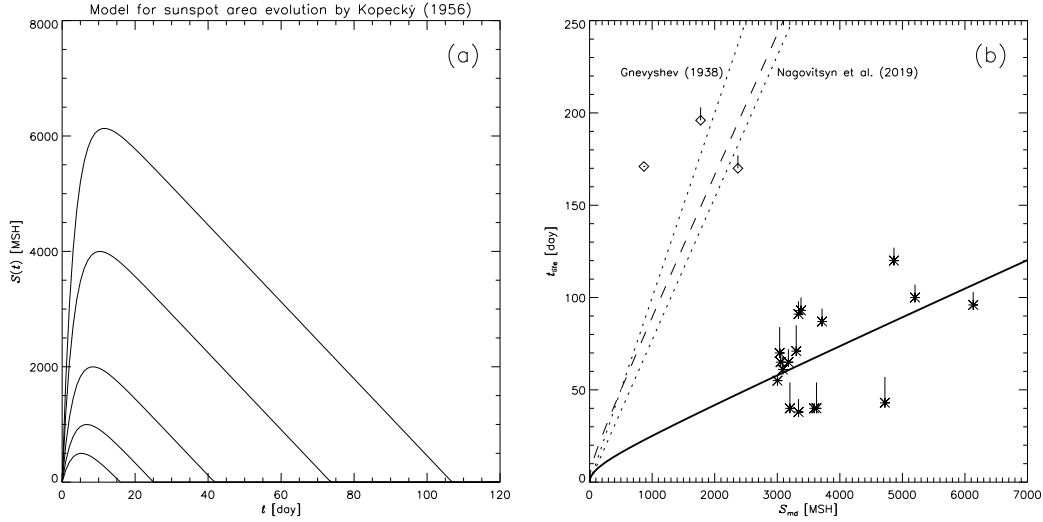}
\end{center}
\caption{
A model for sunspot area evolution proposed by \citet{kop56}. (a) The area $S(t)$ as a function of time $t$ in day is given for a set of parameters $(a, b) = (0.3, 20.0)$, by changing $K$. The curves correspond, from the lowest to the highest, to the maximum-development areas $S_{\rm md}$ of 500, 1000, 2000, 4000, and 6132 MSH.
(b) The lifetime $t_{\rm life}$ as a function of maximum-development area $S_{\rm md}$. The thick solid line is our adopted model with $(a, b) = (0.3, 20.0)$, while the dashed line represents the original relation of \citet{kop56} with $(a, b) = (0.3, 4.0)$. The dotted lines are the formulae given by \citet{gne38} and \citet{nag19}, respectively. The diamond signs and asterisks denote the lifetimes of regions taken from published data \citep{kop84, rgo55}.
}
\label{fig:model}
\end{figure*}

\section{Forward Modeling}
\label{sec:model}
In the analysis so far, we have assumed that the maximum sunspot areas observed on the visible hemisphere are the true maximum areas, which might take place on the back side of the Sun, though. The errors caused by this assumption cannot be corrected, but we may be able to estimate the effects by a forward modeling, namely, by assuming a typical time evolution of sunspot areas we can simulate such effects. As a biproduct, we can convert our maximum-development distribution functions to instantaneous distribution functions.

\begin{figure*}[htbp]
\begin{center}
\includegraphics[width=80mm]{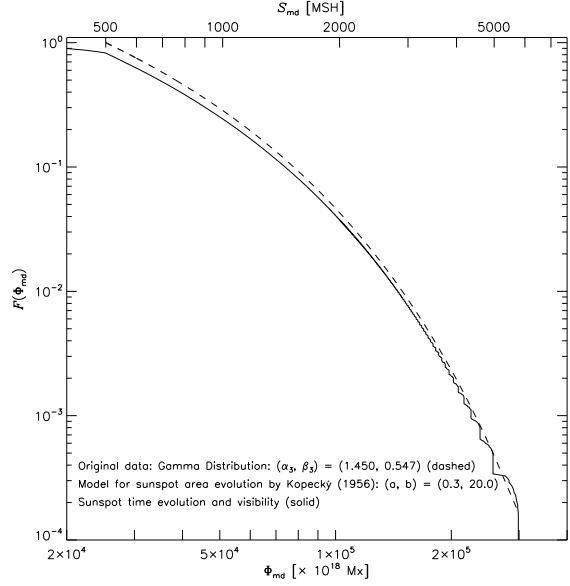}
\end{center}
\caption{
Effects of time evolution of sunspot areas were estimated by using a model by \citet{kop56}. The dashed curve is the model generated from a gamma distribution with $\alpha_3 =$ 1.450 and $\beta_3 =$ 0.547 which reproduces the data well. The solid curve shows the distribution expected from the model by taking into account the visibility probability of regions.
\label{fig:influences}
}
\end{figure*}

\subsection{Model}
\label{sec:modeling}
Many models have been proposed to represent the time evolution of sunspot areas \citep[e.g.][]{kop56,ant86,how92,hat08}. Here we use the model proposed by \citet{kop56} because of its analytical simplicity and versatility. The time evolution of sunspot area $S(t)$ is described by a differential equation
\begin{equation}
\frac{{\rm d}S}{{\rm d}t} = -a S + K - b t \qquad (t \ge 0),
\end{equation}
where $a$, $b$, and $K$ are parameters; $(a, b)$ control the shape of the time profile, and $K$ controls the maximum size of the region. The solution to this equation is given as
\begin{equation}
S = \frac{1}{a} \left[ \left( K + \frac{b}{a} \right) \left( 1 - {\rm e}^{-a t}\right) - b t \right],
\label{eq:model_solution}
\end{equation}
and $S(t)$ takes the maximum value
\begin{equation}
S_{\rm md} = \frac{K}{a} - \frac{b}{a^2} \ln \frac{aK+b}{b}
\end{equation}
at time 
\begin{equation}
t_{\rm md} = \frac{1}{a} \ln \frac{aK+b}{b} .
\end{equation}
$S(t)$ starts from $S(0)=0$ and comes back to $S(t_{\rm life})=0$ where
\begin{equation}
\frac{1- \exp(-a t_{\rm life})}{a t_{\rm life}} = \frac{b}{a K +b} .
\end{equation}
The solution for $t_{\rm life}$, if $a t_{\rm life} \gg 1$, is
\begin{equation}
t_{\rm life} \simeq \frac{K}{b} + \frac{1}{a},
\end{equation}
and if $a t_{\rm life} \ll 1$, it is
\begin{equation}
t_{\rm life} \simeq \frac{2 K}{a K + b} .
\end{equation}
\citet{kop56} used the empirical relations $t_{\rm life} {\rm [days]} \simeq 0.1 S_{\rm md} {\rm [MSH]}$ \citep{gne38} and $t_{\rm life}/t_{\rm md} \simeq 0.094 S_{\rm md} {\rm [MSH]} + 9.3$ \citep{kop53} and adopted $(a, b)=(0.3, 4.0)$ which roughly reproduces these empirical relations for $S_{\rm md} \la 400$ MSH. However, this setting makes the lifetimes of 500 MSH (minimum in our database) and 6132 MSH (maximum) regions as 50 days (1.8 solar rotations) and 480 days (1.3 years, 17.6 solar rotations) respectively, which look too long. \citet{nag19} suggested $t_{\rm life} {\rm [days]} \simeq 0.077 S_{\rm md} {\rm [MSH]}$, a slightly shorter lifetime for the specified $S_{\rm md}$ compared to \citet{gne38}, but this also gives large values of lifetimes. According to \citet{kop84} the longest lifetime of regions in the RGO observations was 8 solar rotations (RGO recurrent series No.~2094, 1970 June 11 -- December 23, 195 days, maximum area = 1774 MSH). The region of largest area (RGO region 14886, 6132 MSH) had a lifetime of 95 days (1947 February 5 -- May 11, observed for 4 rotations). 

As will be discussed in Section \ref{sec:instantaneous}, the gamma function model for $F(\Phi_{\rm md})$ described in Section \ref{sec:expected_life} gives the instantaneous distributions of sunspot area or magnetic flux which are consistent with the observed distributions only if the region lifetimes are much shorter; a reasonable value we found is $(a, b) = (0.3, 20.0)$. The value of $a$ is fixed to 0.3, because the rise time of region growth ($\simeq 1/a$) does not strongly depend on the region size. The lifetimes of the 500 and 6132 MSH regions in this model are 16 and 107 days, respectively. Figure \ref{fig:model}a shows the time profiles $S(t)$ derived for $(a, b) = (0.3, 20.0)$ and for several values of $S_{\rm md}$. Figure \ref{fig:model}b compares the models with $(a, b)=(0.3, 4.0)$ and $(0.3, 20.0)$, and other published results of sunspot lifetimes. The diamond signs denote the data points of three longest-lived regions listed in \citet{kop84}. The asterisks denote the lifetimes of maximum sunspot areas $>3000$ MSH taken from \citet{rgo55}. The regions whose emergence or decay (either of them) were not observed on the visible disk have uncertainties in their lifetimes between 1 and 13 days. Hence we roughly assigned a 7-day error bar. The regions whose emergence and decay (both of them) were not observed on the visible disk were assigned with a 14-day error bar. The model with $(a, b) = (0.3, 20.0)$ goes through the middle of the data points representing large ($> 3000$ MSH) regions.

\begin{figure*}[htbp]
\begin{center}
\includegraphics[width=80mm]{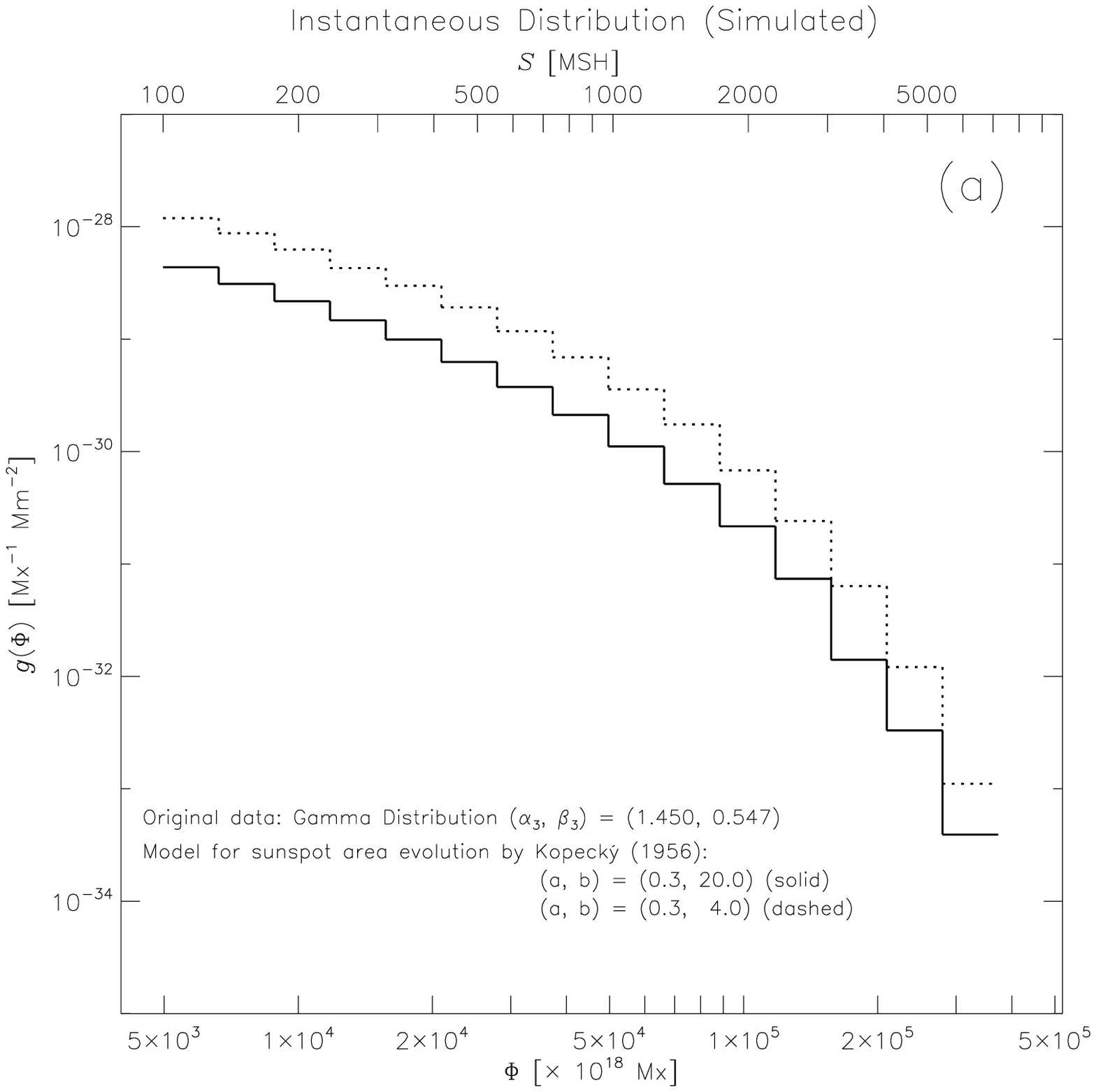}
\includegraphics[width=80mm]{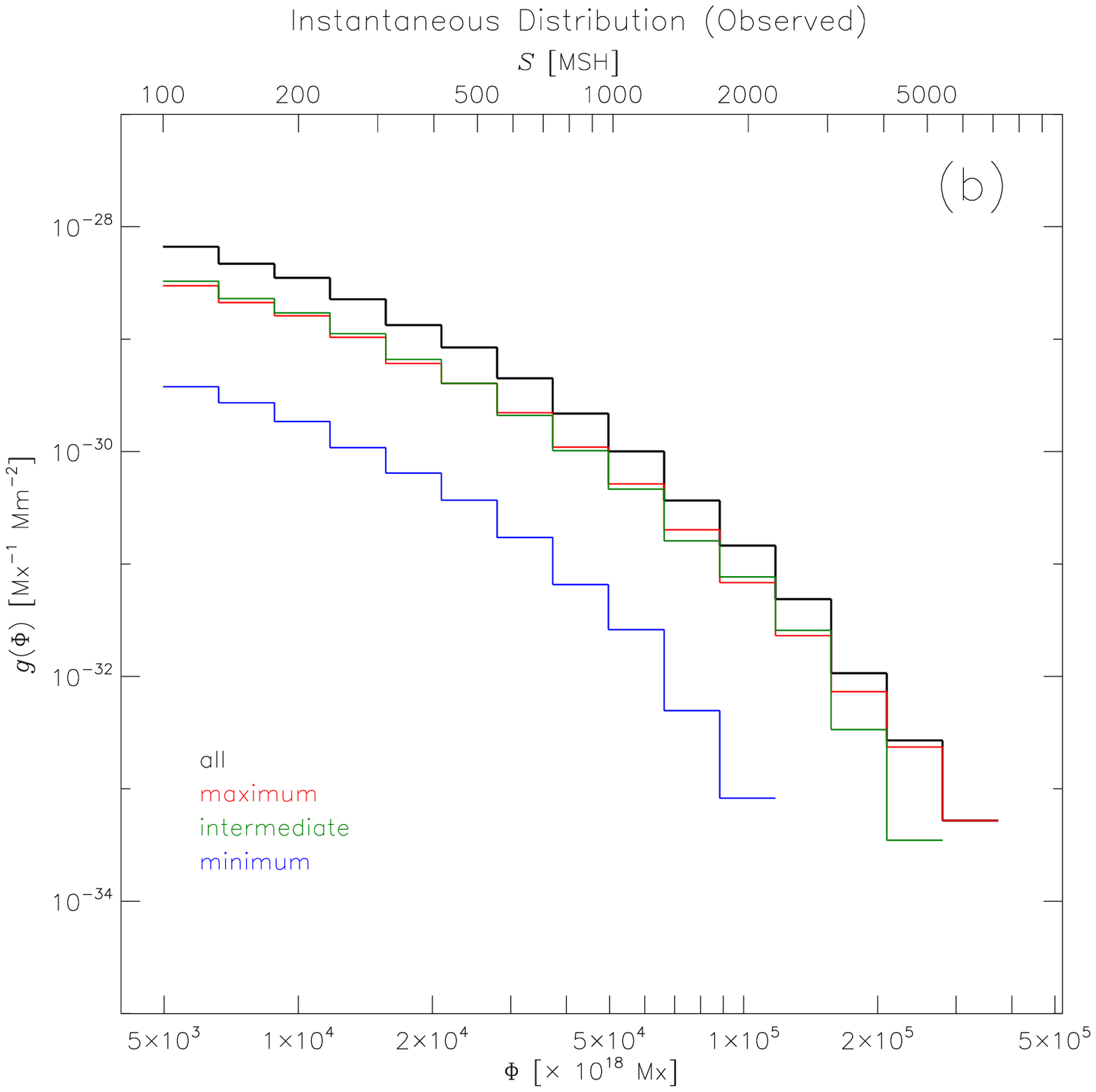}
\end{center}
\caption{
(a) The histogram of the probability distribution function $g(\Phi)$ of instantaneous distribution of magnetic flux, derived from a gamma distribution with $\alpha_3 =$ 1.450 and $\beta_3 =$ 0.547 combined with the sunspot area evolution models of Figure \ref{fig:model}. The solid and dashed histograms correspond to $(a, b)=(0.3, 20.0)$ and $(a, b)=(0.3, 4.0)$, respectively.
(b) The corresponding distributions derived from observations of RGO and NOAA. The histogram in black was derived by using all the data with areas larger than 100 MSH, while red, green, and blue ones are based on data from maximum, intermediate, and minimum phases of activity defined in Table \ref{tab:observation}.
\label{fig:instantaneous}
}
\end{figure*}

\subsection{Effects of Time Evolution}
\label{sec:time_evolution}
The effects of time evolution of sunspots were simulated as follows. By adopting a model of gamma distribution with $\alpha_3 =$ 1.450 and $\beta_3 =$ 0.547, we have generated $N_{\rm s}$ samples ($N_{\rm s}=$ 2995) of $\Phi_{\rm md}$ by
\begin{equation}
\CCDF (\Phi_i) = \frac{i-0.5}{N_{\rm s}} \quad (i=1, 2, \ldots , N_{\rm s}).
\label{eq:generation}
\end{equation}
Then each model was placed at 27 equal-distant longitudes (27 is a rough number of solar rotation period in days), magnetic flux was converted to sunspot area by Equation (\ref{eq:conversion}), and was evolved according to Equation (\ref{eq:model_solution}). The observed maximum values of $S(t)$ in the longitude ranges of $\pm 80^\circ$ were recorded, converted to magnetic flux, and the CCDF was generated.

The solid curve in Figure \ref{fig:influences} shows the result, compared with the true distribution designated by the dashed curve. After fitting the model, we found $\alpha_3 =$ 1.319 and $\beta_3 =$ 0.590. The flattening of the distribution from $\alpha_3 =$ 1.450 to $\alpha_3 =$ 1.319 (by about 0.13) is because we underestimated $S_{\rm md}$, and the data points on the original model (dashed) were shifted toward left on the graph. From this we may estimate that $\alpha_3 =$ 1.450 derived from observations would actually be around $\alpha_3 =$ 1.68, but still our conclusion will hold that the power-law exponent is less than 2.

\begin{table}[htbp]
\begin{center}
\caption{
Summary of observed quantities
\label{tab:observation}
}
\footnotesize
\begin{tabular}{lrrrrrrrr}
\hline
\hline
Data   & \multicolumn{1}{c}{Time} & \multicolumn{1}{c}{Regions per} & \multicolumn{1}{c}{Daily region counts} & \multicolumn{1}{c}{Number of} & \multicolumn{4}{c}{Observed maximum values}\\
\cline{6-9}
source & \multicolumn{1}{c}{span} & \multicolumn{1}{c}{full Sun$^a$} & \multicolumn{1}{c}{per hemisphere} & \multicolumn{1}{c}{observations}  & \multicolumn{1}{c}{Total} & \multicolumn{1}{c}{Total} & \multicolumn{1}{c}{Region} & \multicolumn{1}{c}{Region} \\
       &        & \multicolumn{1}{c}{$N_{\rm s}$} & \multicolumn{1}{c}{$N_{\rm rd}$} & \multicolumn{1}{c}{$N_{\rm d}$} & \multicolumn{1}{c}{area} & \multicolumn{1}{c}{flux} & \multicolumn{1}{c}{lifetime} & \multicolumn{1}{c}{counts} \\
       & \multicolumn{1}{c}{[year]} & \multicolumn{1}{c}{$(S_{\rm md} \ge$} & \multicolumn{1}{c}{($S \ge$} & \multicolumn{1}{c}{($S_{\rm hs} \ge$} & \multicolumn{1}{c}{[MSH]} & \multicolumn{1}{c}{[Mx]} & \multicolumn{1}{c}{[d]} &     \\
       &        & \multicolumn{1}{c}{500 MSH)}      & \multicolumn{1}{c}{500 MSH)}  & \multicolumn{1}{c}{10 MSH)}  &       &      &     &     \\
\hline
All          & 146.71 & 2995  & $1.97 \times 10^4$ & $4.42 \times 10^4$ & 8382$^b$ & $4.36 \times 10^{23}$ & 195$^c$ & 26$^d$ \\
Maximum      &  35.26 & 1466  & $0.99 \times 10^4$ & $1.28 \times 10^4$ & 8382$^{\phantom{b}}$ & $4.36 \times 10^{23}$ &     \\
Minimum      &  37.65 &  114  & $0.06 \times 10^4$ & $0.71 \times 10^4$ & 2268$^{\phantom{b}}$ & $1.16 \times 10^{23}$ &     \\
Intermediate &  73.80 & 1415  & $0.92 \times 10^4$ & $2.43 \times 10^4$ & 8080$^{\phantom{b}}$ & $3.54 \times 10^{23}$ &     \\
\hline
\end{tabular}
\end{center}
$^a$ Recurrent regions were manually picked up and counted only once when they showed the largest area.\\
$^b$ 1947 April 8 \\
$^c$ RGO recurrent series No.~2094, 1970 June 11 -- December 23 \citep{kop84}.\\
$^d$ 1937 July 12 \\
\end{table}

\begin{table}[htbp]
\begin{center}
\caption{
Simulation runs for Sun-as-a-star parameters.
\label{tab:simulation}
}
\small
\begin{tabular}{lrrrrrr}
\hline
\hline
Data         & \multicolumn{1}{c}{Simulation} & \multicolumn{1}{c}{Regions per} & \multicolumn{4}{c}{Simulated maximum values} \\
\cline{4-7} 
source       & \multicolumn{1}{c}{parameters} & \multicolumn{1}{c}{full Sun$^a$} & \multicolumn{1}{c}{Total} & \multicolumn{1}{c}{Total} & \multicolumn{1}{c}{Region} & \multicolumn{1}{c}{Region} \\
             &            & \multicolumn{1}{c}{$N_{\rm s}$} & \multicolumn{1}{c}{area$^b$} & \multicolumn{1}{c}{flux} & \multicolumn{1}{c}{lifetime} & \multicolumn{1}{c}{counts} \\
             &            & \multicolumn{1}{c}{($S_{\rm md} \ge$} & \multicolumn{1}{c}{[MSH]} & \multicolumn{1}{c}{[Mx]} & \multicolumn{1}{c}{[d]} &  \\
             &            & \multicolumn{1}{c}{10 MSH)}  &     &     &     &  \\
\hline
All          &  (0.3, 20.0) & $ 8.5 \times 10^4$       &   8270   & $4.30 \times 10^{23}$ & 101 & 15 \cr
Maximum      &  (0.3, 20.0) & $ 4.2 \times 10^4$       &   7420   & $3.85 \times 10^{23}$ &  93 & 22 \cr
Minimum      &  (0.3, 20.0) & $ 0.3 \times 10^4$       &   3680   & $1.90 \times 10^{23}$ &  65 &  5 \cr
Intermediate &  (0.3, 20.0) & $ 4.0 \times 10^4$       &   6120   & $3.17 \times 10^{23}$ &  93 & 14 \cr
Combined     &  (0.3, 20.0) & $ 8.5 \times 10^4$       &   7420   & $3.85 \times 10^{23}$ &  93 & 22 \cr
Maximum$^c$  &(0.3, \phantom{2}4.0) & $ 4.2 \times 10^4$& 13770   & $7.19 \times 10^{23}$ & 413 & 44 \cr
\hline
\end{tabular}
\end{center}
$^a$ The number of samples in the gamma function model with minimum values extended down to $4.86 \times 10^{20}\ {\rm Mx}$ (10 MSH).\\
$^b$ The numbers do not have significance of four or more digits because of the nature of the simulation. They are arbitrarily rounded off to the nearest ten.\\
$^c$ This is an artificial case of region lifetimes longer than the standard case of $(a, b) = (0.3,20.0)$.
\end{table}

\subsection{Instantaneous Distribution}
\label{sec:instantaneous}
Once we have a model of time evolution of sunspot areas, we can generate the instantaneous distribution function from the distribution of maximum-development areas. First consider a simple case where the time evolution is represented by a step function (i.e. a spot appears with a maximum-development area and stays so until it suddenly disappears). If the lifetimes of regions do not depend on the areas and take a fixed value, the instantaneous distribution function is the same as the maximum-development distribution function. If the lifetimes are proportional to the areas, the instantaneous distribution function would be flatter than the maximum-development distribution function because larger regions live longer and have higher probability of existence in snapshot data. In the general cases where the time evolution of sunspot areas is a function of time like in Equation (\ref{eq:model_solution}), one must resort to numerical simulations to estimate the instantaneous distribution function of sunspot areas or flux emergence rates.

By adopting a model of gamma distribution with $\alpha_3 =$ 1.450 and $\beta_3 = $ 0.547, we have generated $N_{\rm s}$ samples of $\Phi_{\rm md}$ by Equation (\ref{eq:generation}) and converted them to $S_{\rm md}$ by Equation (\ref{eq:conversion}). This time we have extended the lower limit of sunspot area and flux to 100 MSH and $4.97 \times 10^{21}$Mx, so that $N_{\rm s} = 1.8 \times 10^4$; the number of regions with $S_{\rm md} \ge 500$ MSH was still 2995. Next we used Equation (\ref{eq:model_solution}) to evaluate and record daily values of $S(t)$ and $\Phi(t)$, leading to data samples $N_{\rm rd} = 1.90 \times 10^4$ [region $\times$ day per hemisphere] for $S(t) \ge 500$ MSH. In this process we introduced a filter made of 13 consecutive 1's followed by 14 consecutive 0's, mimicking a 27-day modulation of visibility. This filter, replicated many times to cover the lifetimes of regions and its initial point randomly shifted between 0 and 26-th points, was multiplied to daily values of $S(t)$ and $\Phi (t)$. Thus we derived the histogram per unit area (Figure \ref{fig:instantaneous}a, solid histogram) by taking $A$=area of the solar hemisphere=$3.1 \times 10^{6}\ {\rm Mm}^{2}$. Roughly this distribution is fitted by a gamma distribution with $\alpha_3 =$ 0.928 and $\beta_3 =$ 0.654. The power-law exponent decreased (1.450 $\rightarrow$ 0.928), giving a much flatter distribution. The observed value of $N_{\rm rd}$ obtained by counting regions with $S(t) \ge 500$ MSH every day is 1.97 $\times 10^4$ [region $\times$ day per hemisphere] and roughly agrees with the simulated results. On the other hand if we used the evolution parameters $(a, b) = (0.3, 4.0)$, we ended up with a much larger value $N_{\rm rd} =$ 6.10 $\times 10^4$ [region $\times$ day per hemisphere], and the resulting histogram is shown as the dashed line in Figure \ref{fig:instantaneous}a. As a matter of fact we have selected the values $(a, b) = (0.3, 20.0)$ so that the value of $N_{\rm rd}$ roughly matches the observation.

In deriving the instantaneous distribution functions of sunspot magnetic flux from observed data, we have divided the data into three periods; activity maximum, intermediate state, and activity minimum. (The reason why we did so is given in Section \ref{sec:whole-Sun}.) The dates (year and month) of activity maxima/minima are taken from SILSO (Sunspot Index and Long-term Solar Observations)\footnote{https://wwwbis.sidc.be/silso/cyclesminmax}, and we will define the maximum/minimum periods as within 1.356 years from the maximum/minimum dates. The mean lengths of the maximum, intermediate (between maximum and minimum periods), and minimum states are 2.712, 5.424, and 2.712 years, respectively. The mean length of 13 cycles studied here is 10.85 years, and 10.85/8 = 1.356 years. The regions of $N_{\rm s} =$ 2995 with $S_{\rm md} \ge 500$ MSH were divided into maximum (1466), minimum (114), and intermediate (1415) states, respectively (Table \ref{tab:observation}).

Figure \ref{fig:instantaneous}b shows the observed instantaneous distribution functions of sunspot magnetic flux when all the data were used (black histogram) as well as the three activity phases were treated separately (red, green, and blue histograms). We have used all the data with $S \ge 100$ MSH. The histograms are all similar, meaning that they only change the magnitude and not the form of distribution. Particularly the histogram using all the data is reproduced well by the simulated results with $(a, b)=(0.3, 20.0)$ (Figure \ref{fig:instantaneous}a, solid histogram).

\begin{figure*}[htbp]
\begin{center}
\includegraphics[width=80mm]{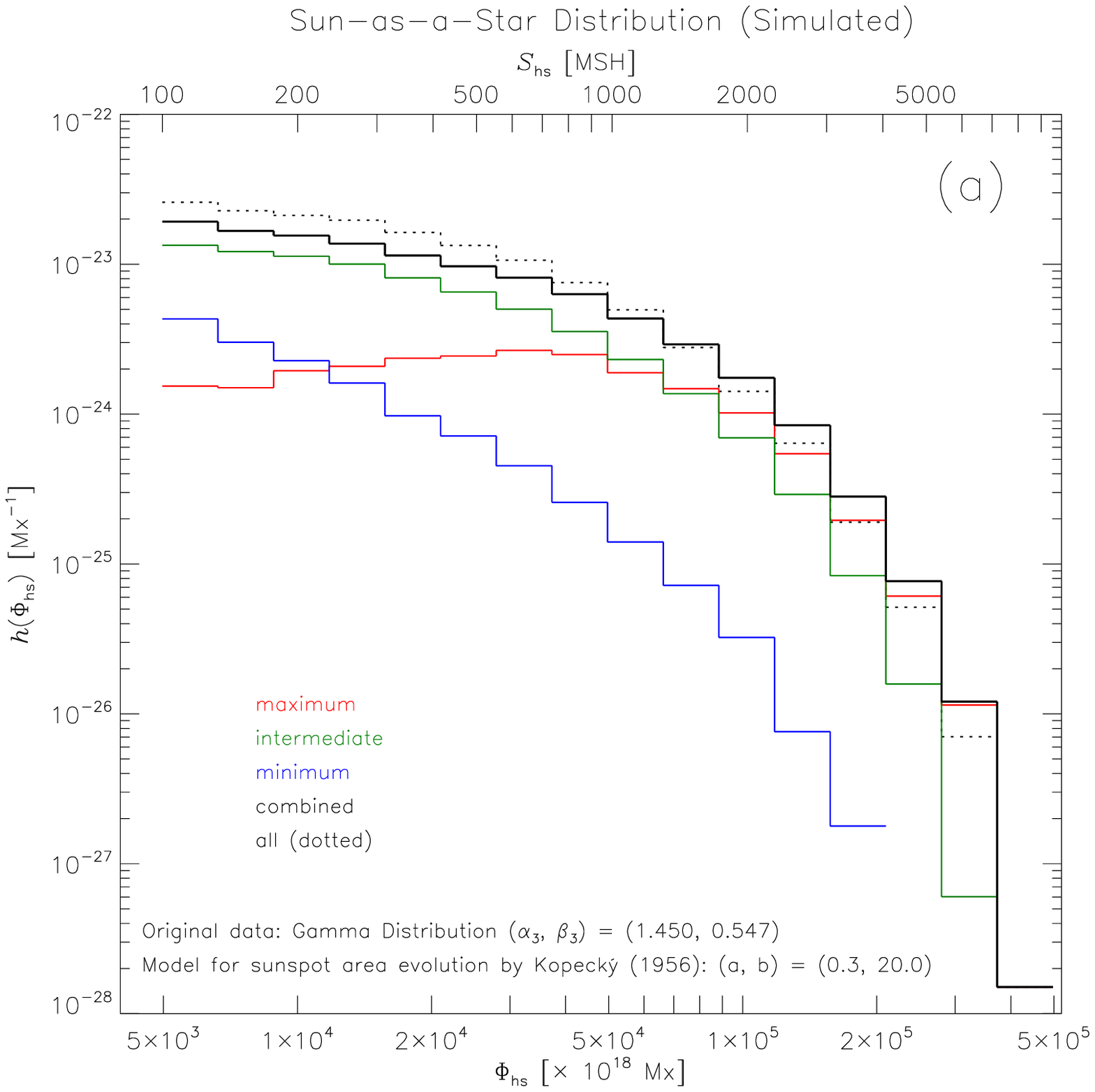}
\includegraphics[width=80mm]{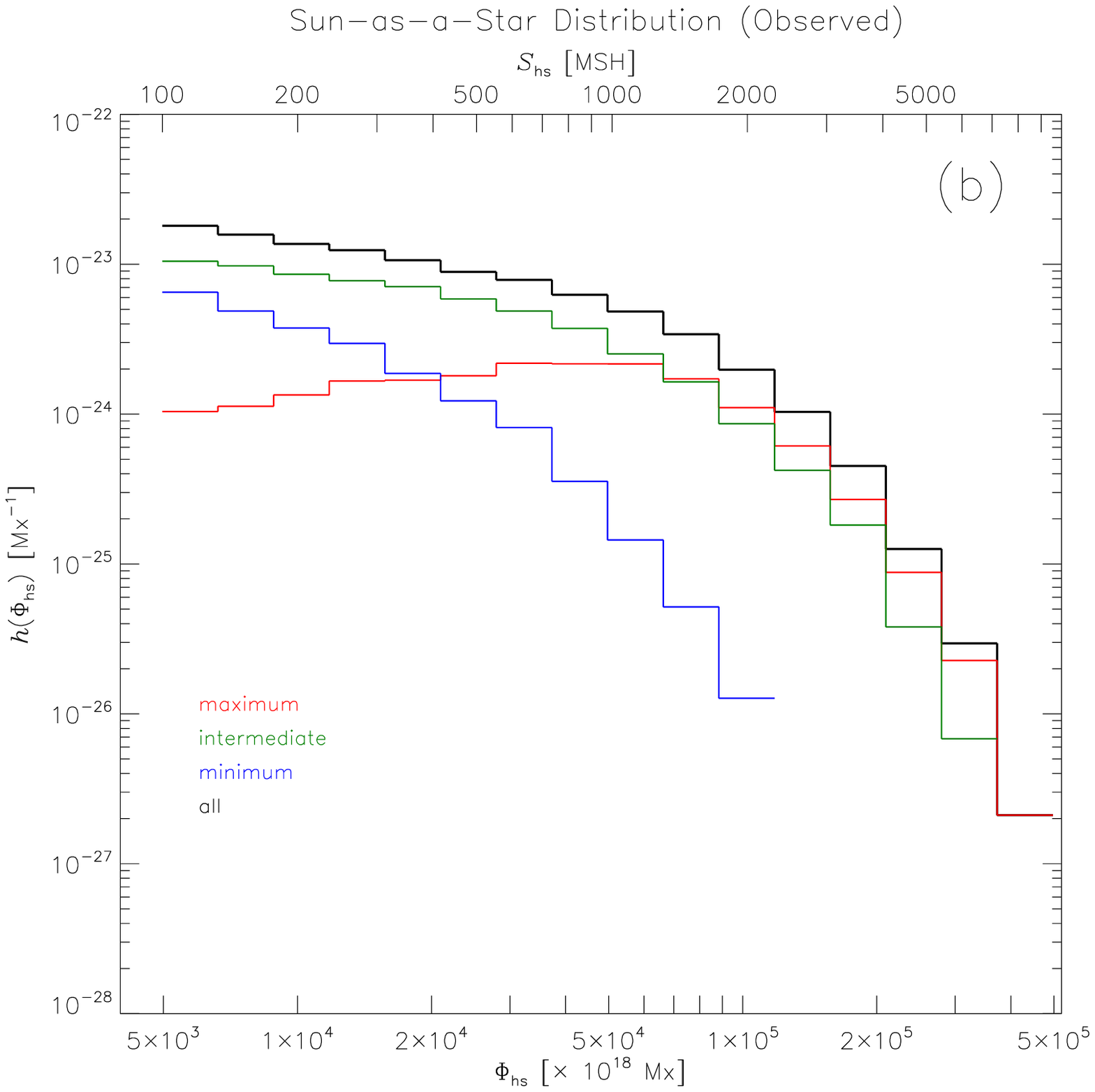}
\end{center}
\caption{
(a) The histograms of the probability distribution function $h(\Phi_{\rm hs})$ of magnetic flux summed over the hemisphere, derived from a gamma distribution with $\alpha_3 =$ 1.450 and $\beta_3 =$ 0.547 combined with the sunspot area evolution model of Figure \ref{fig:model}. The dotted histogram in black was derived by simply using all the generated data. The histograms in red, green, and blue were derived by mimicking activity maximum, intermediate, and minimum states as defined in Table \ref{tab:observation}. The solid histogram in black was made by combining these three histograms.
(b) The corresponding distributions derived from observations of RGO and NOAA. The histogram in black was derived by using all the data with areas larger than 10 MSH, while red, green, and blue ones are based on data from activity maximum, intermediate, and minimum phases of activity defined in Table \ref{tab:observation}.
}
\label{fig:summed}
\end{figure*}

\subsection{Sun-as-a-Star Distributions}
\label{sec:whole-Sun}
From the instantaneous distribution of magnetic flux or sunspot areas, we can also derive the distribution function of total magnetic flux or toral area of sunspot regions, summed over the hemisphere. This is a more fundamental quantity in considering solar irradiance modulation or luminosity variations of stars by starspots. 

By adopting a model of gamma distribution with $\alpha_3 =$ 1.450 and $\beta_3 =$ 0.547, we have generated $N_{\rm s} =$ 2995 samples of $\Phi_{\rm md}$ by Equation (\ref{eq:generation}) and converted them to $S_{\rm md}$ by Equation (\ref{eq:conversion}). The distribution was extended down to $S_{\rm md}=$ 10 MSH, because in the process of time evolution, regions with $S <$ 500 MSH were generated, and these regions also contributed to the total area of sunspots. Then the models were distributed randomly over the time period of $0 \leq t \leq t_0=146.7$ years, and were evolved according to Equation (\ref{eq:model_solution}). If the time evolution of a region hit the end of the time domain $t=t_0$, the profile was folded and continued to $t=0$, to make the statistical distribution stationary in time. The filter of 13 consecutive 1's and 14 consecutive 0's was multiplied to $S(t)$ as in Section \ref{sec:instantaneous}. The daily summed values of magnetic flux over the solar hemisphere, $\Phi_{\rm hs}$, were recorded, and the histogram of PDF was generated (Figure \ref{fig:summed}a). In Figure \ref{fig:summed}, in addition to the abscssa in terms of $\Phi_{\rm hs}$, the scale is also given for the sunspot areas summed over the hemisphere, $S_{\rm hs}$.

Figure \ref{fig:summed}a shows four histograms thus generated. The dotted histogram is the one that was made by simply using all the generated data values. The result is rather counter-intuitive in that the distribution is very flat at the lower end; actually it has a peak at 120 MSH. By looking at the histograms derived from observations (Figure \ref{fig:summed}b), the histogram using all the data is monotonically decreasing, and those derived from activity maximum, intermediate, and minimum states show different behavior. In particular the histogram of the activity maximum period shows a peak at around 3 $\times 10^{22}\ {\rm Mx}$ (650 MSH), but the sum of the three histograms (red, green, and blue) gives the histogram in black which is monotonically decreasing.

The red, green, and blue histograms in Figure \ref{fig:summed}a obtained from simulations mimic the activity maximum, intermediate, and minimum states. If all combined, we obtain the histogram in a black solid line which roughly reproduces the observed histogram in Figure \ref{fig:summed}b (black). Here it must be stressed that the sum of the red, green, and blue histograms in Figure \ref{fig:summed}a does not lead to the black dotted histogram (derived from using all the data) but to the black solid histogram. In the simulated data the solar cycle modulation was not built in and had to be introduced manually. The black dotted histogram (derived from using all simulated data) is similar in shape to the one derived from the intermediate activity state, meaning that the entire simulated data without cycle modulation represent an intermediate activity lasting the entire data period of 146.7 years.

The corresponding histograms (red, green, and blue) in Figures \ref{fig:summed}a and b do not match so well, in various reasons. Our model for the time evolution of sunspot areas is a very idealized one. The lifetime of a region was assumed to be a unique function of maximum-development area, but in reality there should be a statistical distribution of region lifetimes for a given maximum-development area. The division of data into three phases (activity maximum, intermediate, and minimum states) was made by a very simple procedure. Nevertheless the maximum hemispheric summed area of sunspots expected from simulations is about 7420 MSH, which is not very far from the observed maximum value, 8382 MSH. 

\begin{figure*}[htbp]
\begin{center}
\includegraphics[width=80mm]{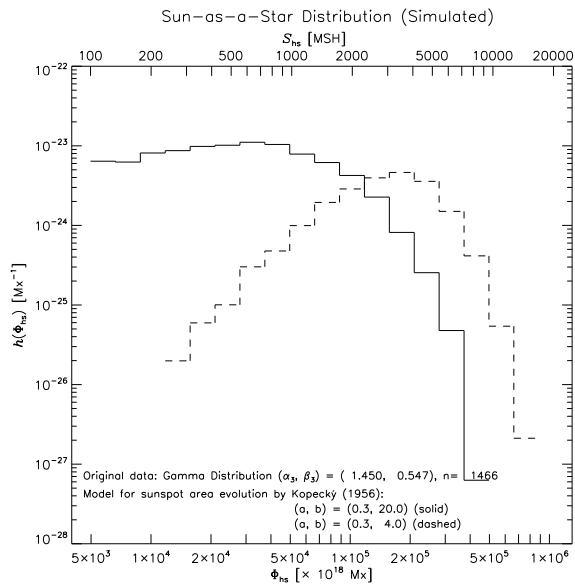}
\end{center}
\caption{
The histograms of the probability distribution function $h(\Phi_{\rm hs})$ of magnetic flux summed over the observed area, derived from a gamma distribution with $\alpha_3 =$ 1.450 and $\beta_3 = $ 0.547 combined with the sunspot area evolution models of Figure \ref{fig:model}. The histogram with a solid line represents the case of our standard setting, $(a, b)=(0.3, 20.0)$. The dashed line shows a case of enhanced lifetimes of sunspots, $(a, b)=(0.3, 4.0)$, which corresponds to the area-lifetime relation suggested by \citet{gne38} for smaller ($< 400$ MSH) sunspots.}
\label{fig:extreme}
\end{figure*}

Figure \ref{fig:extreme} compares two cases of maximum activity phase; one is the same as in Figure \ref{fig:summed} (except for the normalization parameter $T$; Appendix \ref{appendix:normalization}) and the other adopts the parameter setting $(a, b)=(0.3, 4.0)$, i.e. the value originally suggested by \citet{kop56} to reproduce the empirical relations $t_{\rm life} {\rm [days]} \simeq 0.1 S_{\rm md} {\rm [MSH]}$ by \citet{gne38}. The two histograms are not monotonic and have peaks at $\Phi_{\rm hs} = $ 3.4 $\times 10^{22}$ Mx ($S_{\rm hs} =$ 680 MSH) for the former case and at $\Phi_{\rm hs} = 1.8 \times 10^{23}$ Mx ($S_{\rm hs} = 3500$ MSH) for the latter case. The maximum value of the hemispheric summed area for the latter case is13770 MSH. 

The summed area of sunspots can be converted to the modulation in total solar irradiance \citep[TSI, $\simeq$ 1361 W~m$^{-2}$;][]{kopp21} by
\begin{equation}
\frac{\Delta{\rm TSI}}{\rm TSI} = c \frac{S_{\rm h}{\rm [MSH]}}{2 \times 10^6}
\label{eq:TSI}
\end{equation}
with $c \simeq $ 0.22--0.31 \citep{hud82}. If we take the observed maximum value of $S_{\rm hs} = 8382$ MSH (covering 1.67 \% of the visible disk) and $c=0.25$, we obtain $\Delta$TSI/TSI = 1.05 \%. If we assume a hypothetical extreme case of $S_{\rm hs} =$ 13770 MSH discussed above (covering 2.7 \% of the visible disk), we obtain $\Delta$TSI/TSI = 1.7 \%.

The reasons why we revisited the case of $(a, b)=(0.3, 4.0)$ is, on the one hand, to show that this parameter setting leads to overestimated summed total areas of sunspots. On the other hand it is important to remember that the two simulations in Figure \ref{fig:extreme} are based on the same flux emergence rates, or the same strength of dynamo action. Sunspots live longer for the setting of $(a, b)=(0.3, 4.0)$ than the case of $(a, b)=(0.3, 20.0)$ by about a factor of 5, meaning that the former case represents a situation of lower diffusion of sunspot magnetic flux, or possibly weaker surface turbulent convection.

\section{Summary and Discussion}
\label{sec:summary}
We have investigated the probability distributions of sunspot areas $S_{\rm md} \geq 500$ MSH by using the data from RGO (1987 April -- 1976) and USAF/NOAA (1977--2020). Recurrent regions were only counted once at their maximum area development. The data scale of NOAA was adjusted to the scale of RGO by Equation (\ref{eq:calibration}), and the sunspot areas were converted to the magnetic flux contents of active regions by Equation (\ref{eq:conversion}). We obtained a sample of 2995 regions covering 146.7 years.

The data were fitted by a power-law distribution and four two-parameter distributions (tapered power-law, gamma, lognormal, and Weibull distributions). The parameter values were obtained by the MLE method or by minimizing the revised Kolmogorov-Smirnov metric (KSr, Equation (\ref{eq:KSr})). Superiority of one model over the other was assessed by the AIC values. The acceptance or rejection of a specific model was assessed by the conventional Kolmogorov-Smirnov (KS) test, or by using KSr and its P-values (probability of realization).

The power-law model was unfavorable compared to the four models in terms of AIC, and was not acceptable by the classical KS test. Among the two-parameter models, the lognormal and Weibull distributions performed well, but their behavior extended to smaller regions ($S \ll 500$ MSH) did not connect to the previously published results. Therefore, our choices were tapered power-law and gamma distributions. The latter was more favorable considering its consistent sign of curvature of the PDF (Appendix \ref{appendix:tapered-power-law}, Figure \ref{fig:pwrexp}). We also preferred the model determined by KSr=min than the MLE solution; the former directly minimized the deviation of the model CCDF from the observed CCDF, with more weight on the tail of the distribution.

The power-law portion of the tapered power-law and gamma distributions was found to be characterized by its power exponent 1.35--1.9. This was smoothly connected to the power-law distribution of exponent 2 for smaller active regions obtained by \citet{khz93}. Modern observations on ephemeral regions and smaller flux concentrations by \citet{hst03} and \citet{tho11} implied an exponential decline or steeper power law. Although \citet{tho11} suggested that all the magnetic structures were fitted by a single power law with an exponent $\simeq 2.7$, we argue that this is an extreme conclusion. Large sunspots and ordinary (or small) active regions show the power-law behavior with exponent 1.35--2, and their amplitude changes by a factor of 10 between activity maximum and minimum. Small flux concentrations, whether they follow a steeper power law or exponential decline, do not change significantly between activity maximum and minimum, and we suggest that they give way to the active region population at around $\Phi_{\rm md} \simeq 10^{20}$ Mx.

The exponential fall-off of our tapered power-law and gamma distributions were significant, and the expected frequencies of large sunspots (Table \ref{tab:prediction}) were low. The largest sunspot the Sun can generate was estimated to be around $2 \times 10^4$ MSH. Such an upper limit to exist seems reasonable because the magnetic fields generating sunspots must be amplified within the convection zone, which has a finite size of $2 \times 10^5$ km (30 \% of solar radius).

The effects of time evolution of sunspot areas were estimated by introducing a model by \citet{kop56}. Our model assumed the lifetimes of sunspots roughly one fifth of the empirical relation for lifetimes proposed by \citet{gne38}, in order for the flux emergence rates we derived to be consistent with the instantaneous distribution of sunspot areas. Our assumption that the observed maximum area of a region on the visible hemisphere was a reasonable approximation to the true maximum area was confirmed.

By using the same evolution model, we converted the distribution of maximum development areas to the distribution of instantaneous areas, and further derived the Sun-as-a-star distribution functions of total sunspot areas. In the current Sun, the largest total hemispheric areas of sunspots recorded is 8382 MSH, i.e. covering 1.67 \% of the visible disk. By artificially increasing the lifetimes of large sunspots, the total area covered by sunspots can be increased, even up to 2.7 \% of the solar disk, leading to the modulation in TSI of about 1.7 \%.

Recently, \citet{mae17} compared the appearance rates of starspots on slowly rotating solar-type stars and the rates of sunspots and found that the two distributions are not very dissimilar, speculating that the sunspots and larger starspots share a common physical origin. However, in the last 147 years, we observed no sunspots greater than 6132 MSH in area. If the sunspot area distribution were to follow a power law and it were a matter of observation period, perhaps we can find some events of a relevant size by extracting historical records from a period even before that of the RGO data \citep{vaq07}. Our statistical analysis implies that the probability for the Sun to produce much larger single active region is small; 10000 MSH region every 3--8 $\times 10^4$ years. A similar argument on the hemispheric summed areas of sunspots was not made in this paper because we do not have a model for its distribution function, namely how the distribution decays at large values of summed areas. This will be the task in a future paper.

Our hypothetical simulation runs assuming enhanced lifetime of sunspots (not enhanced generation of large sunspots) showed that the total area of sunspots can reach 2.7 \% of the visible disk. The emergence rates of sunspots reflect the process for the solar dynamo to generate magnetic flux, but the lifetime of magnetic structures is controlled by diffusion due to convective eddies. The latter may be an independent process from the dynamo. In the case of the Sun, the diffusive decay of sunspots leads to lifetimes of at most 7 months \citep{kop84}. Under which conditions one may have larger starspots (stronger dynamo) or starspots living for many years (reduced diffusion) could be an important issue in understanding the nature of super-large starspots.

\acknowledgments
The authors would like to thank the referee for useful suggestions that have contributed in improving the data analysis procedures. They also thank Y.~Iida for helpful comments.
The work of Sakurai was supported by JSPS KAKENHI Grant numbers JP15H05816 and JP20K04033.
The work of Toriumi was supported by JSPS KAKENHI Grant numbers JP15H05814, JP16K17671, JP20KK0072, JP21H01124, and JP21H04492, and by the NINS program for cross-disciplinary study (Grant Nos. 01321802 and 01311904) on Turbulence, Transport, and Heating Dynamics in Laboratory and Astrophysical Plasmas: ``SoLaBo-X.''
.

%

\vspace{5mm}
\facilities{SDO(HMI), SOHO(MDI), Hinode(SOT), KPNO}





\appendix

\section{Probability Distribution Functions}	
\subsection{Definitions}
\label{appendix:definition}
To be generic, we will use $x$ instead of $\Phi$ for a statistical variable, and consider a semi-infinite range $\xmin \leq x$. The observed values of $x$ are indicated by $x_i$ ($i=1, 2, \ldots, n$). The probability distribution function (PDF) and its complementary cumulative distribution function (CCDF) for flux emergence rates are denoted by $\PDF (x)$ and $\CCDF (x)$, respectively, which are
\begin{equation}
\PDF (x) = -\frac{{\rm d}\CCDF}{{\rm d} x},
\end{equation}
\begin{equation}
\CCDF (x) = \int_x^\infty \PDF (x')\; {\rm d}x', 
\end{equation}
and
\begin{equation}
\CCDF (\xmin) = \int_{\xmin}^\infty \PDF (x)\; {\rm d}x =1.
\end{equation}
The observed CCDF ($\CCDF_{\rm obs}$) is defined as an aggregate of step functions,
\begin{equation}
\CCDF_{\rm obs}(x) = (\mbox{number of data with~} x_i \geq x)/n.
\label{eq:CCDF_obs}
\end{equation}
At $x=x_i$, $\CCDF_{\rm obs}(x)$ jumps from $i/n$ to $(i-1)/n$, and $\CCDF_{\rm obs}(x_i) = i/n$ because of the ``$\geq$" condition in Equation (\ref{eq:CCDF_obs}).

\subsection{Normalization}
\label{appendix:normalization}
The emergence rate of regions with maximum-development magnetic flux $\Phi_{\rm md}$ is given by Equation (\ref{eq:emergence_rate}),
\begin{equation}
f (\Phi_{\rm md}) = - \frac{N_{\rm s}}{A T} \frac{{\rm d} F(\Phi_{\rm md})}{{\rm d} \Phi_{\rm md}}
\end{equation}
(Figures \ref{fig:data}, \ref{fig:power-law}, \ref{fig:all-models}, \ref{fig:gamma-function}, \ref{fig:comparison}, and \ref{fig:influences}). Here $T=5.36 \times 10^4$ days, $A=6.2 \times 10^6 {\rm Mm}^2$ is the full-Sun area.

For instantaneous distributions of magnetic flux $\Phi$, we use $G(\Phi)$ for CCDF, and the probability distribution $g(\Phi)$ is given by
\begin{equation}
g (\Phi) = -\frac{N_{\rm rd}}{AT} \frac{{\rm d}G(\Phi)}{{\rm d}\Phi}
\end{equation}
(Figure \ref{fig:instantaneous}). Here $T=5.36 \times 10^4$ days, $A=3.1 \times 10^6 {\rm Mm}^2$ is the area of solar hemisphere. $N_{\rm rd}$ is given in Table \ref{tab:observation}.

For the distributions of summed hemispheric magnetic flux $\Phi_{\rm hs}$, we use $H(\Phi_{\rm hs})$ for CCDF and the probability distribution function $h(\Phi_{\rm hs})$ is given by
\begin{equation}
h (\Phi_{\rm hs}) = -\frac{N_{\rm d}}{T} \frac{{\rm d}H(\Phi_{\rm hs})}{{\rm d}\Phi_{\rm hs}}
\end{equation}
(Figures \ref{fig:summed} and \ref{fig:extreme}). Here $T$ is actually the total number of observations (one observation per day), so that $T = 5.36 \times 10^4$ observations in Figure \ref{fig:summed} for all the cases (maximum, intermediate, minimum, all, and combined). In figure \ref{fig:extreme}, $T = 1.3 \times 10^4$ observations. $N_{\rm d}$ is given in Table \ref{tab:observation}.

\subsection{Power-Law Distribution}
\label{appendix:power-law}
\begin{equation}
\CCDF (x) = \left( \frac{x}{\xmin} \right)^{-\alpha_1+1} \qquad (\alpha_1 > 1), 
\end{equation}
\begin{equation}
\PDF (x) = \frac{1}{(\alpha_1 -1) \xmin} \left( \frac{x}{\xmin} \right)^{-\alpha_1}.
\end{equation}
The maximum-likelihood estimator for the power exponent $\alpha_1$ is given by \citep[e.g.,][]{cla09}
\begin{equation}
\alpha_1 = 1 + \left[ \sum_{i=1}^{n} \ln \left( \frac{x_i}{\xmin} \right) \right]^{-1} .
\end{equation}

\subsection{Tapered Power-Law Distribution}
\label{appendix:tapered-power-law}
First we define CCDF as \citep{kag02}
\begin{equation}
\CCDF (x) = \left( \frac{x}{\xmin} \right)^{-(\alpha_2 - 1)}
  \exp \left[ -\beta_2 \frac{x - \xmin}{\xmin} \right]
\qquad (\alpha_2 > 1, \beta_2 \geq 0), 
\end{equation}
($M, M_{\rm t}, M_{\rm cm}$, and $\beta$ in \citet{kag02} are $M=x$, $M_{\rm t}= x_{\rm min}$, $M_{\rm t}/M_{\rm cm}=\beta_2$, and $\beta=\alpha_2$ in our notation) which gives the PDF as follows;
\begin{equation}
\PDF (x) = \frac{1}{\xmin}
\exp \left[ -\beta_2 \frac{x - \xmin}{\xmin} \right]
\left[ \frac{1}{\alpha_2 - 1} \left( \frac{x}{\xmin} \right)^{-\alpha_2} +
\beta_2 \left( \frac{x}{\xmin} \right)^{-(\alpha_2 -1)} \right].
\label{eq:pwrexp_PDF}
\end{equation}
The MLE for parameters $\alpha_2$ and $\beta_2$ are given by \citep{kag01,v-j01}
\begin{equation}
\beta_2 = \left[ n - (\alpha_2 - 1) \sum_{i=1}^{n} \ln \left( \frac{x_i}{\xmin} \right) \right]
\left[ \sum_{i=1}^{n} \frac{x_i-\xmin}{\xmin} \right]^{-1} ,
\end{equation}
\begin{equation}
\sum_{i=1}^{n} \left[ \alpha_2 -1 + \beta_2 \frac{x_i}{\xmin} \right]^{-1}
= \sum_{i=1}^{n} \ln \frac{x_i}{\xmin} .
\label{eq:pwrexp_MLE}
\end{equation}
Equation (\ref{eq:pwrexp_MLE}) must be solved numerically for $\alpha_2$.

\begin{figure*}[htbp]
\begin{center}
\includegraphics[width=70mm]{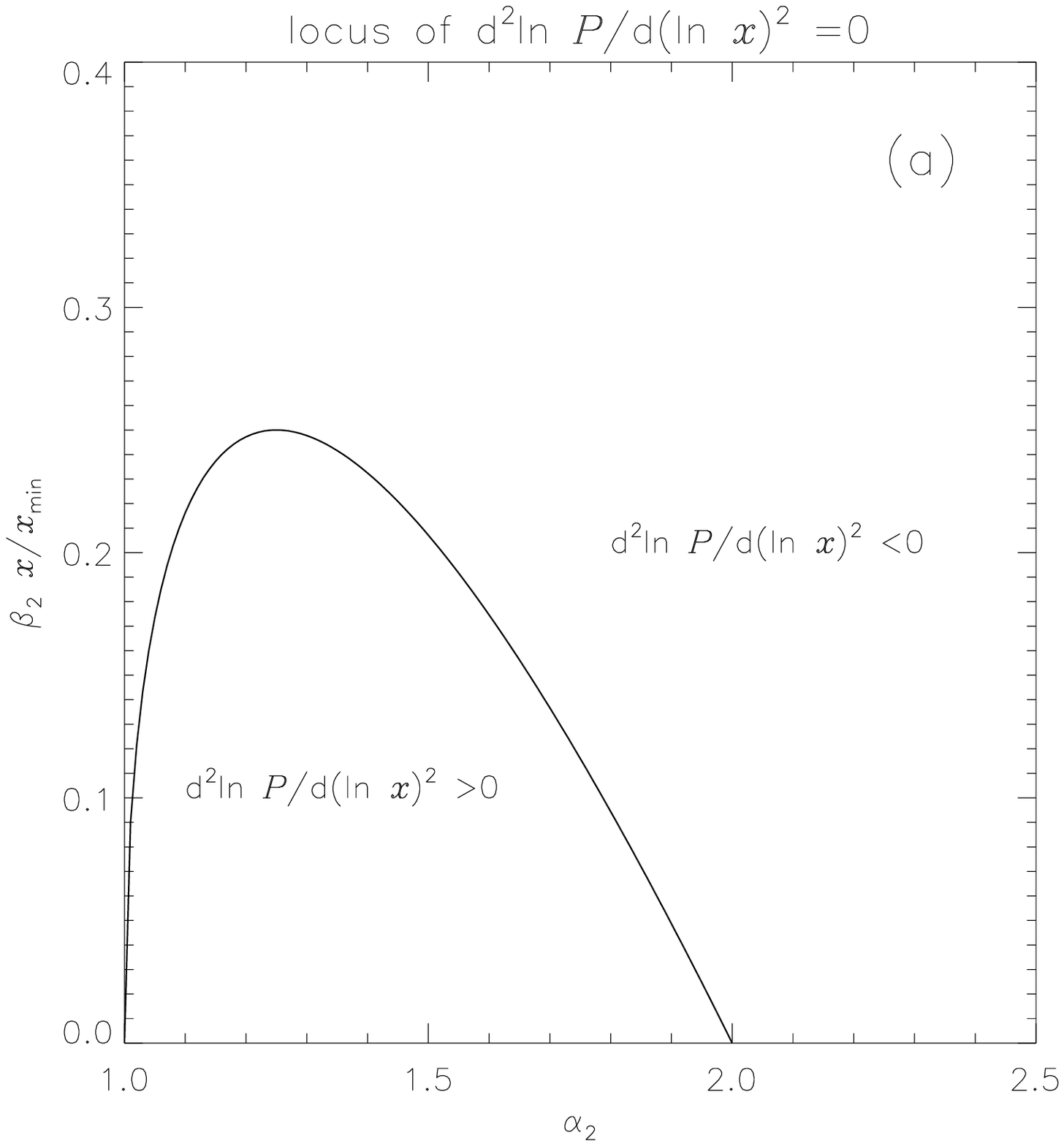}
\includegraphics[width=70mm]{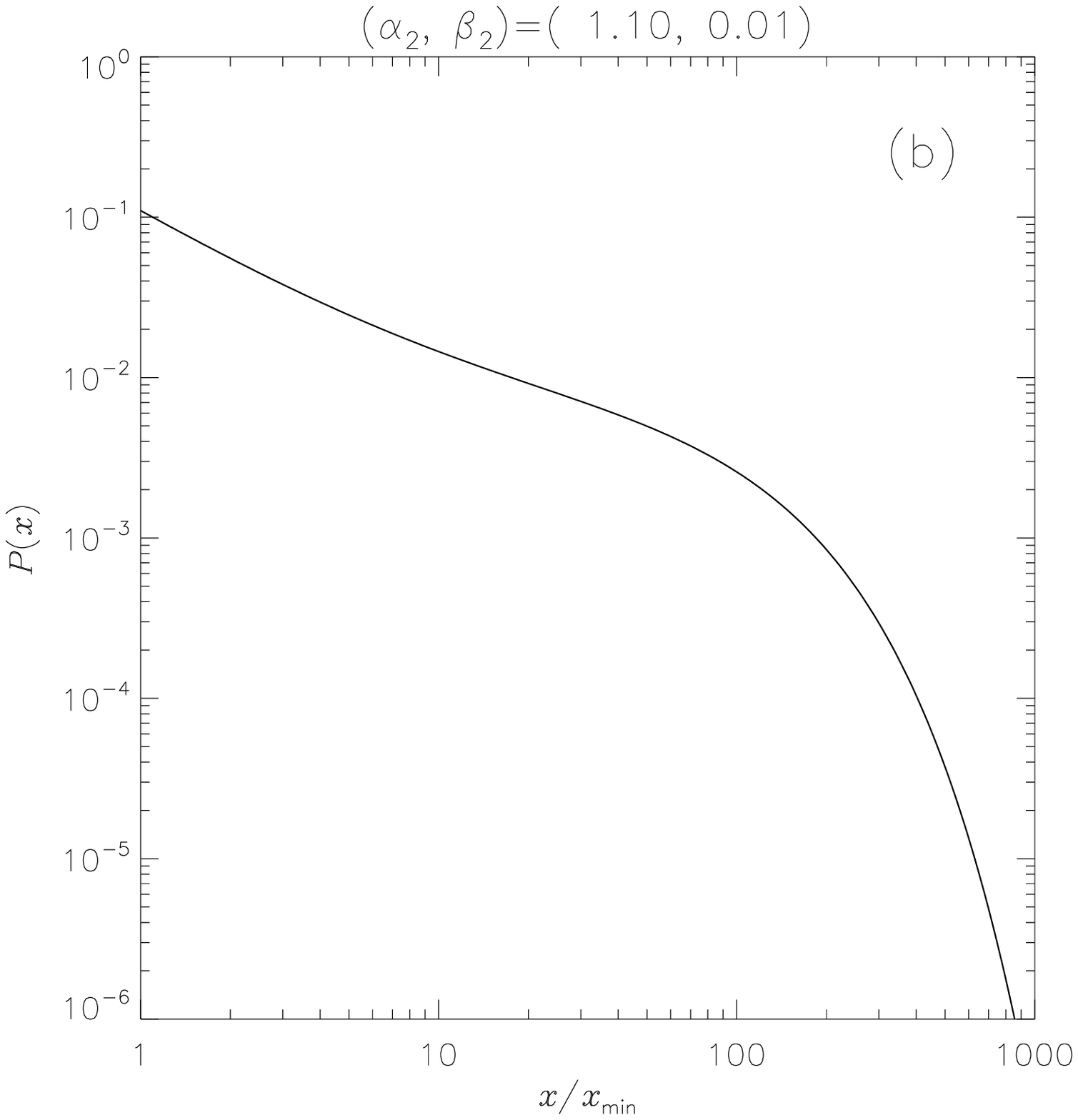}
\end{center}
\caption{
(a) The locus of ${\rm d}^2 \ln \PDF /{\rm d}( \ln x)^2 =0$. In the area above this curve, the slope of $\ln \PDF$ as a function of $\ln x$ monotonically steepens as $x$ increases.
(b) An example of $\PDF (x)$ for $\alpha_2 =1.1$ and $\beta_2 =0.01$. The slope is not monotonic, and an inflection point appears.
\label{fig:pwrexp}
}
\end{figure*}

Equation (\ref{eq:pwrexp_PDF}) indicates that the PDF contains two power-law components with exponents $\alpha_2$ and $\alpha_2 -1$, the former being the dominant component for small $x$. We can show that
\begin{equation}
\frac{{\rm d}^2 \ln \PDF}{{\rm d}( \ln x)^2} 
= - \frac{\beta_2 x}{\xmin} \left[ 1 - \frac{\alpha_2 -1}{(\alpha_2 - 1 + \beta_2 x/\xmin)^2} \right] .
\end{equation}
The sign of this quantity is distributed as in Figure \ref{fig:pwrexp}a. For $\alpha_2 >2$, ${\rm d}^2 \ln \PDF /{\rm d}( \ln x)^2$ is always negative, and the slope of $\ln \PDF$ as a function of $\ln x$ monotonically steepens as $x$ increases. If $1 < \alpha_2 < 2$, ${\rm d}^2 \ln \PDF /{\rm d}( \ln x)^2$ changes sign at $\beta_2 x/\xmin = \sqrt{\alpha_2 -1} - (\alpha_2 -1)$, namely the slope of $\ln \PDF$ as a function of $\ln x$ first flattens and then steepens as $x$ increases. This ``reversed curvature" is conspicuous if $\alpha_2$ is close to 1 (Figure \ref{fig:pwrexp}b), and could be an undesirable feature of this distribution function if the inflection point appears in the fitting range ($x_{\rm min} < x$), or if we extend the distribution down below $x_{\rm min}$.

\subsection{Truncated Gamma Distribution}
\label{appendix:gamma}
We first define PDF as \citep{kag02}
\begin{equation}
\PDF (x) = C \left( \frac{x}{\xmin} \right)^{-\alpha_3}
  \exp \left[ -\beta_3 \frac{x}{\xmin} \right]
\qquad (\alpha_3 > 1, \beta_3 \geq 0), 
\end{equation}
where $C$ is fixed from the normalization condition as
\begin{equation}
C = \frac{ \beta_3^{1-\alpha_3}}{\xmin \Gamma (1- \alpha_3, \beta_3)}
\end{equation}
and $\Gamma$ stands for the incomplete gamma function defined by
\begin{equation}
\Gamma (a, y) = \int_y^\infty t^{a-1} {\rm e}^{-t} {\rm d}t .
\end{equation}
If $\alpha_3 <0$ (which is more common cases in application), $\PDF (x)$ can be defined down to $x=0$. Here we consider the case $\alpha_3 >1$ so that $x$ must be bounded from below as $x \geq \xmin$, hence the name "truncated gamma distribution." CCDF is given by
\begin{equation}
\CCDF (x) = \frac{\Gamma (1-\alpha_3, \beta_3 x/\xmin)}{\Gamma (1- \alpha_3, \beta_3)} .
\end{equation}
In order for $\Gamma$-function not to diverge at $\beta_3 \rightarrow 0$ and $\alpha_3 > 1$, a recurrence formula
\begin{equation}
\Gamma (a +1, y) = a \Gamma (a, y) + y^a{\rm e}^{-y}
\end{equation}
is used to rewrite $\CCDF (x)$ as
\begin{equation}
\CCDF (x) = \frac{%
\Gamma (3-\alpha_3, \eta) -     \eta^{2-\alpha_3}   {\rm e}^{- \eta_3} - (2-\alpha_3)  \eta_3^{1-\alpha_3} {\rm e}^{-\eta}}{
\Gamma (3-\alpha_3, \beta_3) - \beta_3^{2-\alpha_3} {\rm e}^{-\beta_3} - (2-\alpha_3) \beta_3^{1-\alpha_3} {\rm e}^{-\beta_3}}
\qquad (\eta = \beta_3 x/\xmin),
\end{equation}
which can be safely used for $1 < \alpha_3 < 3$.

To obtain the MLE solutions for $\alpha_3$ and $\beta_3$, the log-likelihood LLH,
\begin{equation}
{\rm LLH} = n \ln C - \alpha_3 \sum_{i=1}^{n} \ln \frac{x_i}{\xmin}
- \beta_3 \sum_{i=1}^{n} \frac{x_i - \xmin}{\xmin}
\end{equation}
is usually maximized for both $\alpha_3$ and $\beta_3$ numerically \citep{joh11}. Here we use the condition $\partial$LLH$/\partial\beta_3 =0$ explicitly, i.e.
\begin{equation}
\ln \Gamma (1-\alpha_3, \beta_3) = -\alpha_3 \ln \beta_3 - \beta_3 - \ln \left[
\frac{\alpha_3 -1}{\beta_3} + \frac{1}{n} \sum_{i=1}^{n} \frac{x_i}{\xmin} \right]
\end{equation}
which gives $\beta_3$ if $\alpha_3$ is specified. Therefore, LLH is maximized for $\alpha_3$ to obtain the MLE solution.

\subsection{Truncated Lognormal Distribution}
\label{appendix:lognormal}
We first define PDF as
\begin{equation}
\PDF (x) = \frac{C}{\sqrt{2 \pi} \sigma x}
\exp \left[ -\frac{ \left\{ \ln  (x/\xmin)  - \mu \right\}^2 }{2 \sigma^2} \right]
\label{eq:PDF_lognormal}
\end{equation}
where $C$ is fixed from the normalization condition as
\begin{equation}
C = 2 \left[ 1 - {\rm erf} \left( - \frac{\mu}{\sqrt{2} \sigma} \right) \right]^{-1}
\end{equation}
and erf stands for the error function defined by
\begin{equation}
{\rm erf}(y) = \frac{2}{\sqrt{\pi}} \int_0^y {\rm e}^{-t^2} {\rm d}t .
\end{equation}
The peak of $\PDF$ is at $\xmin \exp (\mu - \sigma^2)$. The lognormal distributions can generally be defined down to $x=0$, and under such cases the mean and the dispersion are given by $\xmin \exp (\mu + \sigma^2/2)$ and $\xmin^2 \exp (2\mu+\sigma^2) [\exp(\sigma^2) -1]$.  In our case we truncate the distribution at $x=x_{\rm min}$, so that the mean and the dispersion are not given by these formula.

CCDF is given by
\begin{equation}
\CCDF (x) = \left[
1 - {\rm erf} \left( \frac{ \ln (x/\xmin) -\mu}{\sqrt{2} \sigma} \right) \right] \left[
1 - {\rm erf} \left( \frac{                -\mu}{\sqrt{2} \sigma} \right) \right]^{-1} .
\end{equation}
By introducing $\zeta = \mu/\sigma$, the MLE solution is given by \citep{cro20}
\begin{equation}
\sigma = \frac{1}{n} \sum_{i=1}^{n} \ln \frac{x_i}{\xmin}
\left[ \zeta + \sqrt{\frac{2}{\pi}} \frac{{\rm e}^{-\zeta^2/2}}{1 - {\rm erf} (-\zeta/\sqrt{2})} \right]^{-1}
\end{equation}
and
\begin{equation}
1 - \sqrt{\frac{2}{\pi}} \zeta  \frac{{\rm e}^{-\zeta^2/2} }{1 - {\rm erf} (-\zeta/\sqrt{2}) } 
= \frac{1}{n} \sum_{i=1}^{n} \left[ \frac{1}{\sigma} \ln \left( \frac{x_i}{\xmin} \right) - \zeta \right]^2 .
\label{eq:lognormal_MLE}
\end{equation}
Equation (\ref{eq:lognormal_MLE}) must be solved numerically for $\zeta$.

\subsection{Truncated Weibull Distribution}
\label{appendix:weibull}
CCDF and PDF are given by \citep{wei39}
\begin{equation}
\CCDF (x) = \exp \left[ -\beta_5 \left( \frac{x}{\xmin} \right)^k + \beta_5 \right]
\qquad (k > 0, \beta_5 >0)
\end{equation}
and
\begin{equation}
\PDF (x) = \frac{\beta_5 k}{\xmin} \left( \frac{x}{\xmin} \right)^{k-1}
\exp \left[ -\beta_5 \left( \frac{x}{\xmin} \right)^k + \beta_5 \right] .
\end{equation}
If $k \ge 1$, $\PDF (x)$ can be defined down to $x=0$. Here we consider the case $k < 1$, so that the equations above are here modified to include the truncation at $x=\xmin$. The MLE solutions are given by \citep{win89}
\begin{equation}
\beta_5 = \left[ \frac{1}{n} \sum_{i=1}^{n} \left( \frac{x_i}{\xmin} \right)^k -1 \right]^{-1}
\end{equation}
and
\begin{equation}
\frac{n}{k} + \sum_{i=1}^{n} \ln \frac{x_i}{\xmin}
\left[ 1 - \beta_5 \left( \frac{x_i}{\xmin} \right)^k \right] =0 .
\label{eq:weibull_MLE}
\end{equation}
Equation (\ref{eq:weibull_MLE}) must be solved numerically for $k$. 

The behavior of $\PDF$ near $x \simeq \xmin$ is
\begin{equation}
\PDF (x) \simeq \frac{\beta_5 k}{\xmin} 
\left[ 1 - \frac{x - \xmin}{\xmin} (1 - k + \beta_5 k) \right] ,
\label{eq:weibull_xmin}
\end{equation}
so that the slope in the $\ln \PDF$ vs.~$\ln x$ plot at $x=\xmin$ is ${\rm d}\ln \PDF/{\rm d}\ln x = -(1-k+\beta_5 k)$.

The behavior of $\PDF$ at $x \simeq 0$ is
\begin{equation}
\PDF (x) \simeq \frac{\beta_5 k}{\xmin} 
\left(\frac{x}{\xmin}\right)^{-(1-k)} \exp(\beta_5)
\label{eq:weibull_x=0}
\end{equation}
so that the distribution approaches a power law with exponent $1-k$.

\section{Evaluation Metrics}	
\label{appendix:metrics}
The goodness-of-fit metrics measure the distance between the theoretical and observed CCDFs. In this paper we use the Kolmogorov-Smirnov metric (KS) and its variant, KSr.

\subsection{Kolmogorov-Smirnov Metric}
\label{appendix:KS}
The KS metric is defined as \citep{ste70,ste16}
\begin{equation}
{\rm KS} = \max_{1 \leq i \leq n} \left( \frac{n+1-i}{n}-\CCDF (x_i), \CCDF (x_i) - \frac{n-i}{n} \right).
\label{eq:KS}
\end{equation}
For large-enough $n$, the probability of obtaining $\sqrt{n}$ KS larger than a specified value is given by the so-called Kolmogorov-Smirnov function with relatively weak dependence on $n$, regardless of the form of CCDF. We used the method proposed by \citet{sim11} to calculate the Kolmogorov-Smirnov function.

\subsection{Modified Kolmogorov-Smirnov Metric}
\label{appendix:KSr}
Since the observed and theoretical CCDFs match at $x \simeq \xmin$ and $x \rightarrow \infty$ by definition, the data points in these regions have less contributions to the KS metric compared to mid data points. As a matter of fact, the contributions scale as $\sqrt{\CCDF (1- \CCDF )}$ \citep{and52}. Therefore, one can obtain more uniform contributions from all the data by dividing the difference by $1/\sqrt{\CCDF (1- \CCDF )}$ \citep{and52, cla09}. In this paper we adopt
\begin{equation}
{\rm KSr} = \max_{1 \leq i \leq n} \left( \frac{n+1-i}{n}-\CCDF (x_i), \CCDF (x_i) - \frac{n-i}{n} \right) /
\sqrt{\CCDF (x_i)},
\label{eq:KSr}
\end{equation}
to emphasize the tail portion only. A mismatching near $x \simeq \xmin$ may occur from physical (e.g. limitation in detecting small sunspots, which is very unlikely in the present study because 500 MSH sunspots are big) or technical reasons. Therefore, we do not want to amplify it.

\section{Comparison with Published Data}	
\label{appendix:comparison}
Here we will summarize the flux emergence rates [Equation (\ref{eq:emergence_rate})] given in the literature and how they were converted and plotted in Figure \ref{fig:comparison} in units of Mx$^{-1}$ Mm$^{-2}$ d$^{-1}$. The values of $n$, $A$, and $T$ differ in individual data sources.

\subsection{\citet{tho11}}
\label{appendix:tho11}
\citet{tho11} analyzed the emergence rates of small-scale magnetic field patches using the data from Hinode/SOT and obtained a formula extending all the way up to the active-region scales,
\begin{equation}
f_{\rm ThP} = \frac{n_0}{\Phi_0} \left( \frac{\Phi}{\Phi_0} \right)^{-\alpha} \qquad [\mbox{Mx$^{-1}$ cm$^{-2}$ day$^{-1}$}]
\end{equation}
where $n_0 = 3.14 \times 10^{-14}$ cm$^{-2}$ day$^{-1}$, $\Phi_0 = 1.0 \times 10^{16}$ Mx, and $\alpha= -2.69$. In Figure (\ref{fig:comparison}) is plotted $\tilde{f}_{\rm ThP} = f_{\rm ThP} \times 10^{16}$, the last factor is (Mm/cm)$^2$.

Figure 5 of \citet{tho11} were also copied to Figure \ref{fig:comparison} using geometrical approximations to their curves.

\subsection{\citet{khz93}, \citet{sch94}}
\label{appendix:khz93}
\citet{khz93} derived the emergence rates of bipolar active regions using the data obtained at NSO Kitt Peak \citep{liv76}. The data were taken on 739 days nearly uniformly distributed from 1975 to 1986. Table \ref{table:KHZ} reproduces the data give in their Table 1 before various corrections (data gaps, etc.) were applied. We have converted the area data $A_j$ and $\Delta A_j$ in square degrees (sq.~deg. = $1.48 \times 10^{18}$ cm$^2$) to magnetic flux as
\begin{eqnarray}
\Phi_j       \mbox{[Mx]} &=& 150 \times        A_j \times 1.48 \times 10^{18} , \\
\Delta\Phi_j \mbox{[Mx]} &=& 150 \times \Delta A_j \times 1.48 \times 10^{18} .
\end{eqnarray}
The conversion factor 150 Mx cm$^{-2}$ (or mean field strength) was taken from \citet{sch94}, who suggested 136 and 153 derived from different methods. The histogram in Figure \ref{fig:comparison} shows the counts $N_j /(\tau_{\rm obs} \Delta\Phi_j S_{\rm h})$ where $\tau_{\rm obs} = 739$ d and $S_{\rm h} = 3.04 \times 10^6$ [Mm$^2$] is the area of solar hemisphere.

\citet{sch94} reported that the area distribution $N_j(A_j)$ is fitted by a power law with exponent $p=2$, namely
\begin{equation}
\tilde{N}_j(A_j) = a^\ast A_j^{-p}, 
\end{equation}
with amplitudes $a^\ast = 1.23$ for activity minima and $a^\ast = 10$ for activity maxima. This range of values is shown in Figure \ref{fig:comparison} as two parallel lines, after the following conversion
\begin{equation}
\tilde{N}(\Phi_j) = a^\ast A_j^{-p} / [150 \times (1.48 \times 10^{18}) \times S_{\rm h}] .
\end{equation}

\begin{table}
\begin{center}
\caption{
Data reproduced from \citet{khz93}. Here $j$ represents the bin number, $A_j$ and $\Delta A_j$ represent the area and bin with in square degrees, and $N_j$ represents the number of regions of the $j$-th bin.
}
\begin{tabular}{lrrrrrrrrrrr}
\hline
\hline
$j$ &           1 &    2&    3&    4&    5&    6&    7&    8&    9&   10&   11 \cr
\hline
$A_j$ &       2.90& 3.93& 4.95& 5.95& 6.99& 8.00& 8.91& 9.92&11.34&13.13&15.47 \cr
$\Delta A_j$ & 1.0&  1.0&  1.0&  1.0&  1.0&  1.0&  1.0&  1.0&  2.0&  2.0& 2.0 \cr
$N_j$ &        313&  155&   77&   65&   44&   32&   30&   24&   56&   26&  27 \cr
\hline
\hline
$j$ &           12&   13&   14&   15&   16&   17&   18&   19&   20&   21&  22 \cr
\hline
$A_j$ &      17.29&19.32&22.63&26.15&30.10&35.89&44.07&51.66&60.70&68.50&73.30 \cr
$\Delta A_j$ & 2.0&  2.0&  4.0&  4.0&  4.0&  8.0&  8.0&  8.0&  8.0&  8.0& 8.0 \cr
$N_j$ &         30&   17&   23&   17&    9&   23&    3&    5&    1&    0&   1 \cr
\hline
\hline
\end{tabular}
\end{center}
\label{table:KHZ}
\end{table}

\subsection{\citet{hst03}}
\label{appendix:hst03}
\citet{hst03} investigated the emergence rates of small-scale bipolar magnetic field patches (ephemeral regions) using the data from SOHO/MDI. The data were taken between 1996 and 2001. In their Figure 11 they presented the flux emergence rates in number of regions per day per $10^{18}$ Mx bin over the whole solar surface for the data of 1997 October and 2000 August as
\begin{eqnarray}
f_{\rm H} &=& S_{\rm cm} (28.4 \times 10^{-20}) \exp(- \Phi/(10^{18} {\rm [Mx]} \times 5.5 f_{\rm c}) )	\qquad \mbox{(1997 October)} , \\
f_{\rm H} &=& S_{\rm cm} (21.1 \times 10^{-20}) \exp(- \Phi/(10^{18} {\rm [Mx]} \times 5.2 f_{\rm c}) )	\qquad \mbox{(2000 August)} .
\end{eqnarray}
Here $1 \times 10^{18}$ Mx $ < \Phi \lesssim 6 \times 10^{18}$ Mx, $S_{\rm cm} = 6.1\times 10^{22}$ is the whole solar area in cm$^2$, and $f_{\rm c}= 1.6$ is a conversion factor to put the MDI field strength to the scale of NSO Kitt Peak. In Figure \ref{fig:comparison} these were converted to
\begin{equation}
\tilde{f}_{\rm H} = f_{\rm H} / (1.0 \times 10^{18} \mbox{[Mx]} \times S_{\rm Mm})
\end{equation}
where $S_{\rm Mm} = 6.1 \times 10^6$ is the whole solar area in Mm$^2$.

\subsection{\citet{bau05}}
\label{appendix:bau05}
\citet{bau05} presented the lognormal distributions fitted to the RGO sunspot area data. They adopted the form
\begin{equation}
\ln \left( \frac{{\rm d}N}{{\rm d}A} \right) = - \frac{ (\ln A - \ln \langle A \rangle )^2}{2 \ln \sigma_A}
+ \ln \left( \frac{{\rm d}N}{{\rm d}A} \right)_{\rm max}
\end{equation}
with the normalization
\begin{equation}
\int_{A_{\rm min}}^\infty \frac{{\rm d}N}{{\rm d}A} {\rm d}A = 1 
\end{equation}
and parameter values $A_{\rm min} = 60$ MSH, $\langle A \rangle = 62.2$ MSH, and $\sigma_A = 2.45$ MSH. These are converted to our standard form [Cf. Equation (\ref{eq:PDF_lognormal})]
\begin{equation}
\PDF_{\rm B} (A) = \frac{\tilde{C}}{\sqrt{2 \pi} \tilde{\sigma} A}
\exp \left[ -\frac{ \left\{ \ln  (A/A_{\rm min})  - \tilde{\mu} \right\}^2 }{2 \tilde{\sigma}^2} \right]
\end{equation}
by taking $\tilde{\mu} = \ln \langle A \rangle + \ln \sigma_A$, $\tilde{\sigma} = \sqrt{\ln \sigma_A}$. Then, by applying Equation (\ref{eq:conversion}), we obtain Equation (\ref{eq:PDF_lognormal}) with $\mu=\tilde{\mu} p$, $\sigma = \tilde{\sigma} p$ ($p=1.02)$.

\subsection{\citet{gop18}}
\label{appendix:gop18}
\citet{gop18} made a general study of the applicability of Weibull distribution to various indices of solar activity. He used the formula for CCDF, represented by $y$, as
\begin{equation}
\log y = a \left[ 1 - \exp \left( -\frac{\gamma - \log x}{\eta} \right) \right] .
\end{equation}
When $x$ is the sunspot area in MSH observed at RGO and NOAA (1874 May to 2016 December) and $y$ is the number of regions with area $>x$, \citet{gop18} gave the parameter values $a=2.5$, $\gamma =3.3$, and $\eta = 0.8$. This should match our representation for CCDF,
\begin{equation}
\CCDF_{\rm G} = n \exp \left[ \beta - \beta \left( \frac{x}{\xmin} \right)^k \right] 
\qquad ( x \geq \xmin ).
\end{equation}
Here we adopt the un-normalized CCDF with a factor $n = 41433$ representing the total number of data. The parameters are related as follows.
\begin{eqnarray}
k &=& \left[ \eta \ln 10 \right]^{-1} , \\
\beta &=& a \ln 10 - \ln n , \\
\xmin &=& 10^{\gamma} \left( 1 - \frac{1}{a} \ln n \right)^{1/k} .
\end{eqnarray}
After converting sunspot area $x$ to magnetic flux $\Phi$ by Equation (\ref{eq:conversion}), and using $t_{\rm obs} = 5.21 \times 10^4$ d (from 1874 May to 2014 December) and $S_{\rm obs} = 6.1 \times 10^6$ (the whole solar area in Mm$^2$), a Weibull PDF multiplied by $n/(t_{\rm obs} S_{\rm obs})$ is plotted in Figure \ref{fig:comparison}. We assume that \citet{gop18} did not make particular considerations on recurrent regions, so that some mismatching is anticipated.

\end{document}